\documentclass[nonacm, sigconf]{acmart}
\usepackage{amsmath}
\usepackage{algorithm}
\usepackage{algpseudocode}

\def\*#1{\mathbf{#1}}
\newcommand*{\expe}{\mathbb{E}}
\newcommand*{\real}{\mathbb{R}}
\newcommand*{\kl}{\mathbb{KL}}

\begin{document}
\settopmatter{printacmref=false}

\title{ExCalibR: Expected Calibration of  Recommendations}

\author{Pannagadatta Shivaswamy}
\email{pshivaswamy@netflix.com}
\affiliation{%
  \institution{Netflix Inc.}
  \city{Los Gatos}
  \state{CA}
  \country{USA}
  \postcode{95032}
}


\begin{abstract}
In many recommender systems and search problems, presenting a well balanced set of results can be an important goal in addition to serving highly relevant content. For example, in a movie recommendation system, it may be helpful to achieve a certain balance of different genres, likewise, it may be important to balance between  highly popular versus highly personalized shows. Such balances could be thought across many categories and may be required for enhanced user experience, business considerations, fairness objectives etc.  In this paper, we consider the problem of calibrating with respect to any given categories over items. We propose a way to balance a trade-off between relevance and calibration via a Linear Programming optimization problem where we learn a doubly stochastic matrix to achieve  optimal balance in expectation.  We then realize the learned policy using the  Birkhoff-von Neumann decomposition of a doubly stochastic matrix. Several optimizations are considered over the proposed basic approach to make it fast. The experiments show that the proposed formulation can achieve a much better trade-off compared to many other baselines. This paper does not prescribe the exact categories to calibrate over (such as genres) universally for applications. This is likely dependent on the particular task or business objective. The main contribution of the paper is that it proposes a framework that can be applied to a variety of problems and demonstrates the efficacy of the proposed method using a few  use-cases.

\end{abstract}

\maketitle

\section{Introduction}
\label{sec:intro}
Many real-world recommender systems optimize their presentation of results for relevance to their users (e.g. \cite{prodrec, youtube16, Gomez-Uribe2015,Agarwal2014}). A popular way of doing this is simply by training a prediction model and then sorting results based on the scores from the trained model. More recently, there is increased emphasis on achieving other types of objectives than just relevance--for example, diversity \cite{diversity-survey}, fairness of different kinds \cite{fairness-survey, ml-survey2}, calibration \cite{Steck18,mip,confidence-aware} etc. In this paper, we consider the calibration framework in \cite{Steck18} and show that we can achieve much better trade-offs between calibration and relevance by an exact optimization rather than a greedy procedure.

In \cite{Steck18}, the author argues that calibration can be a useful goal for recommendation along with relevance. As a motivating example, if a user gives some signal from their usage pattern that they are interested in certain genres such as drama or romance and especially movies on Friday evenings, the recommendation could be calibrated to satisfy their needs.
Similarly, if users from a particular country enjoy local content, and the underlying global model is not recommending enough local content, the recommendation could be calibrated to balance local and non-local content.  Likewise, in search problems, there could be multiple intents behind a search query. From prior interactions of users with such a query, we may have prior knowledge about the ratio of different intents. It may be a useful goal to also preserve a desired distribution over different intents in the context of a search query along with the relevance of search results. 

While we give the above examples (recommendations, search ) as motivation, the basic framework developed in this paper is applicable to many different scenarios.  We assume that along with items (such as movies, search results, products etc.), we also have {\em calibration categories}  for every item (such as genres, local content or not, which interests are covered by a search result, seller identity etc.). We propose a way to calibrate a ranked list or slate of recommendations towards a specified baseline. The baseline itself can be specified in various ways depending on the needs of particular applications. In this paper, we explore three different baselines, but the framework itself is not limited to the presented use-cases and can easily go beyond the use-cases studied in this paper.

% Birkhoff-von Neumann decomposition has been studied extensively in some recent literature e.g., \cite{SinghJoachims2018, WangJoachims2021,Rijke2022} they were mostly to ensure fairness of different kinds in terms of exposure. In particular, some of these papers propose giving exposure proportional to the perceived merit of an item. In contrast, in this paper, we clearly focus on a genre based calibration as originally defined in \cite{Steck18}. Compared to exposure to utility ratio based notion of fairness, we directly trade-off between the relevance score to deviation in the distribution between historic data and the ranking policy we are learning. Thus, although there are some similarities with the prior work, the exact framework, measuring the difference between two distributions and the trade-offs used in this paper differ from existing literature.

The main contributions of this paper are as follows. We start with a formal setup for the calibration problem for the given calibration categories. While most of the previous work focused on set selection calibration problem \cite{Steck19,mip,confidence-aware}, our focus is on a ranked-list or a slate level optimization problem. With our setup, we show that calibration attribute distribution can be described in terms of known quantities and an unknown doubly stochastic matrix that we seek to learn. The  learning objective is set up in such a way that we trade-off between the relevance of the ranking and the calibration of categories with respect to a target that can be specified based on particular requirements.  Compared to the greedy formulations proposed earlier \cite{Steck18} our formulation solves a Linear Program and achieves a better trade-off between relevance and deviation between achieved attribute distribution and ideal attribute distribution.  We then use Birkhoff-von Neumann decomposition to derive policies that achieve the desired trade-offs in expectation.  We show that many additional optimizations can be done to solve the optimization problem fast.    In our experiments, we show that the proposed approach can achieve better trade-off between relevance and calibration compared to greedy approaches and even previously proposed mixed integer programming approaches. While many search or recommender systems can have tens of thousands to millions of items, the end user typically sees a handful of items presented to them on the User Interface on the web or on a mobile phone. The framework in this paper is about getting those top few items well calibrated and generally works with a small candidate set. 

\section{Related Work}
\label{sec:related}

Calibration is a topic that has been studied for long in the context of machine learning (e.g.  \cite{calibrated-bayes, foster}). Historically, the goal in these types of works is to ensure that estimated probability scores from models were well calibrated according to some definitions. More recently, there is a renewed interest in calibration \cite{Steck18,calibration-popularity-bias20, Mesut, mip, confidence-aware} in the context of recommender systems problems.  The work from \cite{Steck18} is the direct framework we use and compare against in this paper where the author considers a trade-off between relevance and a divergence measure between genre (which is used as the calibration category) achieved versus genre distribution desired. In \cite{mip} the authors study a mixed integer programming approach for calibration building upon the framework of \cite{Steck19}. In \cite{confidence-aware}, the authors make some progress towards having a different level of calibration depending on the number of observations of the users. In \cite{calibration-popularity-bias20} the authors study the connection between popularity bias and how it can lead to (mis)calibration.  In \cite{Mesut}, the authors explore the relationship between calibrated recommendation and intent aware recommendation. A closely related concept is diversity in recommendations \cite{diversity-survey}. There are two types of diversity studied in the literature, intrinsic diversity \cite{intrinsic-div} where the goal is to cover as many sub-topics for a query versus extrinsic diversity \cite{extrinsic-div} where the goal is to hedge against uncertainty in the intent.  There are also several works that optimize diversity in recommendation and search problems \cite{greedy-diversity, search-diversity,diversity-recommender}. The author in  \cite{Steck18} argues that calibration as defined in that work can be used for enforcing diversity of one kind by choosing an appropriate prior. 

There is considerable interest in the topic of fairness and bias in machine learning and recommender systems recently \cite{fairness-survey, ml-survey2}.  One line of work focuses on trading off relevance and fairness measures directly in the model learning objective in various ways \cite{pmlr-v80-agarwal18a,Singh/Joachims/19a, pmlr-v80-hashimoto18a, blackbox}.  In contrast, the work in this paper and several other works \cite{WangJoachims2021, Rijke2022, SinghJoachims2018} are focused on post-processing ranking once a model has already been trained. The post-processing approaches can be complement to any improvements in model training itself since it is typically possible to change the underlying model with an existing post-processing technique.

Fairness of exposure is a running theme in many papers that have been using Birkhoff-von-Neuman decomposition \cite{WangJoachims2021, Rijke2022, SinghJoachims2018}. A notion of fairness of exposure was introduced in \cite{SinghJoachims2018} and the authors consider optimizing a ratio of exposure to utility at item level. Subsequently, in  \cite{WangJoachims2021} consider a two-sided market place problem where they proposed an approach to achieve user-fairness, item-fairness and diversity of ranking.   For example, fairness of exposure at item level. In \cite{Rijke2022} considers the problem of fairness of exposure items when the exposure to an item cannot be properly estimated due to inter item dependencies. A main concept emerging out of these works is to be fair in terms of the ratio of exposure to utility.  In contrast, in this paper, we consider a more soft trade-off between the relevance achieved along with a measure of divergence between a prior category distribution and the category distribution induced by a ranking as a way to define calibration. 
\section{Preliminaries}
\label{sec:prelim}
Bold face mathematical symbols represent vectors or matrices. We use $\*1$ ($\*0$) to denote a vector of all ones (zeros) with its dimensionality inferred from the context. For vectors or matrices $\*x$, $\*y$, $\*x \ge \*y$  is an element-wise greater than (or equal to) operation. Similarly, $\*x \ge 0$ denotes all elements of $\*x$ being non-negative. Such notations are used for both matrices and vectors.

We assume that there are $n$ candidate items for a user (e.g. movies, music, products etc.) scored by a ranking system (e.g. a deep-neural network). For simplicity of exposition, we first assume that we need to re-rank all the candidates. We will relax this assumption later. In many recommender systems,  such a ranking is typically done by scoring all $n$ candidate items using, for example, a deep-neural network and then sorting them by their score. The vector $\mathbf{s} \in \real^n$ denotes the scores of all $n$ items with the score of the $j^\text{th}$ item being denoted by $\mathbf{s}_j$. These scores could be personalized or could be a global score per item depending on the application.

In many applications, items  ranked at the top position get much more exposure and attention from users compared to lower positions \cite{position-bias}. We denote by $\*e \in \real^n$ the position-values of the $n$ positions in the ranking. Here, $\*e_j$ the weight of the $j^\text{th}$ position in the ranking. We assume that such position-values are known apriori. Estimating such values has been studied independently (e.g., \cite{position-bias-1}) and is beyond the scope of this paper. As simple examples, motivated from Normalized Discounted Cumulative Gain (NDCG) \cite{ndcg}, we can set $\*e_j \propto \frac{1}{\log(j + 1)}$  for the $j^\text{th}$ slot in the ranking. We assume that the weights $\*e_j$ are non negative and add to one, i.e., $\*e \ge 0$ and $\*e^\top \*1 = 1$. We are often interested in the top $k$ items in the ranking rather than re-ranking all $n$ items. In this case, we simply assume that the position value of $\*e_j = 0$ for all $ j > k$ and $\*e_j > 0$ for all $j \le k$. The position weight vector $\*e$ is still assumed to be non-negative and  that the weights add to one.

Next we assume that there are $r$ possible {\em calibration categories}.  Every candidate item $i$ is endowed with its own distribution over the categories. We consider a general setup where for candidate item $i$ can belong to one or more categories. We denote by $\*A \in \real^{n \times r}$ the matrix of  membership of all items to all categories.  $0 \le \*A_{ij} \le 1$ denotes category membership of $i^\text{th}$ item  to $j^\text{th}$ category. For any given item $i$, the category memberships are assumed to be such that  $\*A \*1 = \*1$. This assumption means that an item can belong to multiple categories, but over all such categories, the membership values add to one. For example, if an item belongs to {\em Action} and {\em Drama} $\*A_{ij}$ could be $0.5$ for $j's$ that correspond to those two genres and $0$ for other genres. Thus, for any item $i$, the $i^\text{th}$ row can be viewed as specifying a full distribution over all  categories. 

The work in this paper relies on learning a so-called Doubly Stochastic Matrix. A doubly stochastic matrix has non-negative values, each row sums to one and each column adds to one as well. Formally, a  matrix $\*P \in \real^{n \times n}$ is doubly stochastic if $\*P_{ij} \ge 0$, $\*P \*1 = \*1$ and $\*P^\top \*1 = \*1$. We denote by $\mathcal{DS}(n)$ the family of all doubly stochastic matrices of size $n \times n$, formally defined as follows:
\begin{align}
    \label{eq:ds-family}
 &\mathcal{DS}(n):=  \\
 &\{ \*P \in \real^{n \times n} | \*P_{ij} \ge 0,~~ \forall 1 \le i,j \le n,~~ \*P\*1 = \*1,~~ \*P^\top \*1 = \*1\}.  \nonumber 
\end{align}

Given $n$ items and $n$ positions, $\*P_{ij}$ of a doubly stochastic matrix can be seen as a way of expressing the probability of placing the item $i$ at position $j$, with the rows representing the items and the columns the positions. Since any row adds to one, a row defines the probability with which a given document $i$ ends up at each of the $n$ positions. Similarly, since each column adds to one, we get the probability of any document ending at position $j$.

A specialized version of doubly stochastic matrices are called permutation matrices. Like a doubly stochastic matrix, each row and each column adds to one in a permutation matrix. In addition, each element of a permutation matrix is either zero or one. This means that every row and every column has exactly one element set to one and all other elements set to zero. We define the set of all permutation matrices of size $n \times n$ as follows:
\begin{align}
    \label{eq:perm-family}
 & \mathcal{P}(n) := \\
 & \{\*P \in \real^{n \times n} | \*P_{ij} \in \{0,1\},~~ \forall 1 \le i,j \le n,~~ \*P \*1 = \*1,~~ \*P^\top \*1 = \*1\}.   \nonumber
\end{align}
A permutation matrix gives a deterministic way to re-order items to positions. Given a permutation matrix $\*P$, if $\*P_{ij} = 1$, it can be seen as assigning the $i^\text{th}$ item to $j^\text{th}$ position. Next, we discuss how a doubly stochastic matrix can be  expressed as a convex combination of permutation matrices. We summarize various notations introduced in this section in Table~\ref{tab:my_label} for quick reference.

\begin{table}[]
    \centering
    \begin{tabular}{|c|c|c|}
    \hline
    Notation & Short Description & Key Properties \\
    \hline
    $n$ & item count &   \\
    $k$ & slot count & $k <=n$ \\
    $r$ & calibration category count &    \\
    $\*e \in \real^n$ & normalized position weights & $\*e \ge \*0, \*e^\top \*1 = 1$ \\
    $\*s \in \real^n$ & item scores &  \\
    $\*A \in \real^{n \times r}$ &  category distribution over items & $\*A \ge \*0, \*A\*1 = \*1$ \\
    \hline
    \end{tabular}
    \caption{Various notations used throughout this paper.} 
    \vskip -0.2in
    \label{tab:my_label}
\end{table}

\subsection{Birkhoff-von Neumann (BVN) decomposition}
In this paper, we learn doubly stochastic matrices that can achieve trade-offs between relevance and deviation from a baseline distribution. Suppose  we have a doubly stochastic matrix $P \in \mathcal{DS}(n)$, the following lemma allows a way to realize a doubly stochastic matrix in terms of permutation matrices \cite{WangJoachims2021, Rijke2022, SinghJoachims2018} known as the Birkhoff-von Neumann (BVN) decomposition:
\begin{lemma}
Any doubly stochastic matrix $\*P \in 
\mathcal{DS}(n)$ can be decomposed into a convex combination of $l=\mathcal{O}(n^2)$ permutation matrices, i.e.,
\begin{align}
    \*P = \sum_{i=1}^l \theta_i \*Q^i
\end{align}
where $\theta_i \ge 0$, $\sum_{i=1}^l \theta_i = 1$ and $\*Q^i \in \mathcal{P}(n)$.
\end{lemma}

The actual algorithm for achieving the above decomposition simply constructs an adjacency matrix from a given doubly stochastic matrix by setting any non-zero entry to 1. A maximum bipartite matching algorithm is then solved to obtain a matching of rows to columns or in other words a permutation matrix. The coefficient for this permutation matrix is set to minimum value in the original doubly stochastic matrix where the permutation matrix has a non-zero value. The permutation matrix is then subtracted from the original doubly stochastic matrix with a multiplicative coefficient set to the coefficient. The procedure is continued until the original doubly stochastic matrix reduces to all zeros. The details of this algorithm can be found in many recent references \cite{WangJoachims2021, Rijke2022, SinghJoachims2018}.

\begin{algorithm}[thp]
\caption{${BVN}(\*P)$}
\label{bvn-policy}
\begin{algorithmic}[1]
    \State Input: $ \*P\in \mathcal{DS}(n)$ 
    \State Output: $\*Q \in \mathcal{P}(n) $ 
    \State Decompose $\*P = \sum_{j=1}^l \theta_i \*Q^i,~ \*Q^i \in \mathcal{P}(n)$ 
    \State $\*Q = \*Q^i$ with probability $\theta_i$ 
    \State {\bf return } $\*Q$
\end{algorithmic}
\end{algorithm}

Given a doubly stochastic matrix, we first decompose it into a convex combination of permutation matrices and then return the permutation matrix $Q_i$ with probability $\theta_i$. This policy is outlined in Algorithm \ref{bvn-policy}. For $Q$ given by BVN($\*P$), by the linearity of expectation, we can write down the following lemma:
\begin{lemma}
\label{lem:expectation}
For the $\theta$ realized through BVN($\*P$), we have, for any $\*x$ and $\*y$ of appropriate dimensions,
\begin{align}
    \*P &= \expe_{\*Q \sim BVN(\*P)} [\*Q] \nonumber \\ 
    \*P \*x &= \expe_{\*Q \sim BVN(\*P)} [\*Q \*x] \nonumber \\ 
    \*y^\top\*P \*x &= \expe_{\*Q \sim BVN(\*P)} [\*y^\top \*Q \*x]. \nonumber
\end{align}
\end{lemma}

Thus, the above decomposition provides a convenient way of going from a doubly stochastic matrix to a realizable policy to rank items with known mean quantities. In the rest of the paper, we focus on learning a doubly stochastic matrix for ensuring calibration along with relevance.
The above expectation equalities provide a way to control the mean behavior of the ranking policy.

The BVN decomposition can be obtained in time ${\mathcal O}(n^4\sqrt{n})$ where $n$ is the number of rows and columns in the doubly stochastic matrix $\*P$ \cite{matching1} where the complexity is dominated by that of the bipartite matching algorithm ${\mathcal O}(n^2\sqrt{n})$. The overall time complexity of the decomposition seems daunting at first. However, this problem is solved only at a small scale on the top few items that are going to be presented. It turns out the approach can be solved nearly as fast as a simple greedy implementation with tools available publicly. 
\section{Calibrated Recommendations}
In this section, using the notations introduced in Section \ref{sec:prelim}, we propose an exact optimization to solve a calibrated recommendation problem. We first describe the greedy approach of \cite{Steck18}.

We follow the setup of \cite{Steck18} and assume that there is a base distribution over categories such as genres $\*q\in \mathbf{R}^r$ where we remind that this prior could be dependent on the user. Many distributions are possible for this prior. For example, as outlined in \cite{Steck18}, $\*q$ could be the distribution over prior view history of the user. For a user with no prior view history, this could be a uniform or another predetermined distribution over all calibration categories. Similarly, the more recent plays could be weighted more in this prior distribution. We simply assume that this distribution is available.

In \cite{Steck18}, the author consider a set selection problem where the goal was to select a set of items ${\mathcal I}$ such that the sum of the scores of the selected were as high as possible while the difference between the distribution over categories induced by the set of items selected was as small as possible with respect to the baseline distribution $\*q$.
Formally, \cite{Steck18} considered the following objective to maximize the trade-off between the scores of the items and the KL divergence between the two distributions:
\begin{align}
\label{eq:kl-greedy}
    (1 - \lambda) \sum_{i \in \mathcal{I}} s_i - \lambda \kl(\*q|\*p(\mathcal{I}))
\end{align}

In the above equation $\kl(\*q|\*p(\mathcal{I}))$ denotes the Kullback-Leilbler (KL) divergence between $\*q$ and $\*p(\mathcal{I})$. The KL divergence between two distributions $\*q$ and a distribution $\*p$ is defined as follows:
\begin{align}
  \kl(\*q|\*p) = \sum_{i=1}^r \*q_i \left( \log(\*q_i) - \log(\*p_i)\right) 
\end{align}

Further, in (\eqref{eq:kl-greedy}), $0 \le \lambda \le 1$ is a parameter that trades-off between the relevance score and the KL divergence. The goal of using the KL divergence is a way to keep the distribution induced by a particular selection of items and the baseline distribution as close to one another as possible. Next, we propose our proposed framework for calibration.

%Next, we consider the above optimization problem \eqref{eq:kl-greedy} as a guideline and propose a position-weighted ranking optimization rather than a sub-set selection problem. We propose linear programming problem that can maximize the relevance  score while keeping the distribution induced by the ranking close to the baseline distribution. While we do not directly make use of KL divergence in our optimization, the goal still is to keep the two distributions $\*p$ and $\*q$ close to one another.

\subsection{Expected Calibration Formulations}
Given a ranking of the items with a doubly stochastic matrix $\*P \in \mathcal{ DS}(n)$, we first need to determine the resulting distribution over the calibration categories from this ranking. We recall from Section \ref{sec:prelim} that $\*A$ is a matrix denoting the distribution of categories over items. We define the following quantities for any $\*P \in \mathcal{ DS}(n)$ and any $\*Q \in \mathcal{P}(n)$.

\begin{align}
\label{eq:genre-distribution}
\*p(\*P) := \*A^\top \*P \*e \text{~~~~~and~~~~~}
\*p(\*Q) := \*A^\top \*Q \*e.  
\end{align}

\begin{lemma}
\label{lemma:prob}
 $\*p(\*P)$ denotes a valid distribution over categories. In other words, $\*p(\*P) \ge \mathbf{0}$ and $\*p(\*P)^\top \*1 = \*1$.
%Moreover, 
%\begin{align}
%\label{sec:distribution}
%\*q(\*P) =  \*A^\top \*P \*e    = \expe_{\*Q \sim BVN(\*P) }[ \*q(\*Q)]
%\end{align}
\end{lemma}
\begin{proof}
We first consider $\*p(\*Q)$ and show that it satisfies the properties of a distribution as well.
First, non-negativity of $\*p(\*Q)$ follows from the non-negativity of $\*A$, $\*Q$ and $\*e$. Next, we consider,
\begin{align}
    &\*1^\top \*p(\*Q)  = \*1^\top \*A^\top \*Q \*e    
  = (\*A \*1)^\top \*Q \*e 
  = (\*A \*1)^\top \Tilde{e} 
  = \*1^\top\Tilde{e}  = 1. \nonumber
\end{align}
In the first equation, we simply used the definition of $\*p(\*Q)$. In second equality, we moved the transpose outside the product. In the third equality, since $\*Q$ is a permutation, we replaced $\*e$ with a reordered version $\mathbf{\tilde{e}}$. Further, from the definition of how the calibration category matrix was defined, we used the equality $\*A\*1=\*1$ (refer to Section~\ref{sec:prelim} and Table~\ref{tab:my_label}). Finally, from the definition of $\*e$ the sum over the individual elements  of $\*e$ in any order is equal to one.

We now take the expectation over $\*Q \sim BVN(\*P)$ which immediately gives (from the expectation lemma~\ref{lem:expectation}):
\begin{align}
    &\*1^\top \*p(\*P) = 
    \expe_{\*Q \sim BVN(\*P) }[\*1^\top \*p(\*Q)]   =1. \nonumber 
\end{align}    
Since $\*p(\*P) =\expe_{\*Q \sim BVN(\*P) }[\*p(\*Q)]$, non-negativity of $\*p(\*P)$ also follows immediately since it is an expectation over individual $\*p(\*Q)$ that themselves are non-negative in each element. 
\end{proof}

We next look at  $\*p(\*P) = \*A^\top \*P \*e$ to get some intuition. $\*P \*e$ denotes the expected amount of exposure that each item gets. Since $\*A^\top$ represents the distribution of categories over items, $\*p(\*P)$ denotes the expected amount of exposure that each calibration category gets in turn. As we saw from Lemma \ref{lemma:prob}, it represents the distribution induced by a particular ranking or from a particular doubly stochastic matrix. We should thus aim to keep $\*p(\*P)$ close to the baseline distribution while achieving good ranking performance. We now propose the following optimization problem that we call {\bf ExCalibR}:

\begin{align}
\label{eq:excalibr-abs}
\max_{\*P, \epsilon \ge 0}~ & (1 - \lambda) \mathbf{s}^\top \*P \*e - \lambda \epsilon^\top \*1 \\
\text{s.t.}~~& \*A^\top \*P \*e - \*q \le \epsilon  \nonumber \\
&  \*q - \*A^\top \*P \*e \le \epsilon   \nonumber \\
& \*P \in \mathcal{DS}(n) \nonumber
\end{align}

The objective of the above optimization problem is linear in $\*P$ and $\epsilon$, the constraints are also linear in $P$ and $\epsilon$. The constraint that $\*P \in \mathcal{DS}(n)$ which ensures double stochasticity can also be posed as three sets of linear constraints as shown in the definition of doubly stochastic matrices. The above optimization can be solved exactly using any LP solver. The constraints ensure that the expected value of the induced distribution is not too far from the baseline distribution. The above formulation \eqref{eq:excalibr-abs} has $n^2 + r$ variables and $2r + n^2 + 2n$ constraints $n^2 + 2n$ constraints are from ensuring that $\*P$ is a doubly stochastic matrix (non-negativity constraints, rows sum to one constraints and columns sum to one constraints).  $r$ which is the number of categories and can  typically be much smaller than $n^2$. Other variants of ExCalibR such as bounding the relative difference between the realized distribution over categories and the baseline distribution, specifying an epsilon and minimizing the deviation using non-negative slack variable etc. are possible. We leave these as potential future research directions.

\subsection{Position Weighted Greedy Procedure}

In equation  \eqref{eq:kl-greedy} we considered a greedy formulation with no weights on different positions as a set optimization problem. To make the comparison fair with our weighted version of ExCalibR, we also considered a weighted version of KL-divergence based weighted objective in a straight-forward way motivated from \eqref{eq:kl-greedy}. Denoting by $\mathcal{I}$ with ordered items $j_1, j_2, \ldots, j_{|\mathcal{I}|}$, where $j_i$ denotes the item added at position $i$ of the ranking, we have 

\begin{align}
\label{eq:kl-greedy-weighted}
    \max_{\mathcal{I}} ( 
     (1 - \lambda) \sum_{i=1}^{|\mathcal{I}|} \*s_{j_i} \*e_i
   - \lambda \kl(\*q|\*p(\mathcal{I})) ).
\end{align}
In the above $\*p(\mathcal{I})$ denotes the distribution over categories induced by position-weighted ranking given by the ordered set $\mathcal{I}$.  With the above formulation, the relevance part (first term) is similar to that in equation (\ref{eq:excalibr-abs}) except that we are learning a doubly stochastic matrix whereas the greedy optimization is simply learning one re-ordering. The second part of the objective is KL divergence similar to that in \eqref{eq:kl-greedy}. We optimize the above objective akin to the greedy optimization in (\ref{eq:kl-greedy}). We initially start with an empty list. We add the element that maximizes the above objective the most given the current list and so on until we select an element for all slots of recommendation. When the size of $\mathcal{I}$ is equal to $n$, the above objective can be expressed concisely in our notation as learning an ordering (a permutation matrix $\*Q$) as follows: 

\begin{align}
\label{eq:kl-greedy-weighted2}
    \max_{\*Q \in \mathcal{P}(n) } \left( 
     (1 - \lambda) \mathbf{s}^\top \*Q \*e
   - \lambda \kl(\*q|\*p(\*Q)) \right).
\end{align}
However, directly solving the above is a hard problem, hence the greedy objective above \eqref{eq:kl-greedy-weighted} gives a practical way to solve it. One might also ask why not try to directly optimize an objective like the one in equation (\ref{eq:kl-greedy-weighted}) instead of the LP formulation we defined in (\ref{eq:excalibr-abs})? However, we note that if we consider 
$-\kl(\*q|\*p(\*P))$ in our maximization objective,
\begin{align}
  -\kl(\*q|\*p(\*P)) &= -\sum_{i=1}^r \*q_i \log\left(\frac{\*q_i}{\*p_i(\*P)}\right)  \nonumber \\
  &= -\sum_{i=1}^r \*q_i \log(\*q_i) + \sum_{i=1}^r \*q_i \log(\*p_i(\*P)) \nonumber \\
  & = -\sum_{i=1}^r \*q_i \log(\*q_i) + \sum_{i=1}^r \*q_i \log(\*A_{:i}^\top \*P \*e) \nonumber \\
 & = -\sum_{i=1}^r \*q_i \log(\*q_i) + \sum_{i=1}^r \*q_i \log(\*A_{:i}^\top \sum_{j=1}^m \theta_j \*Q^j \*e) \nonumber \\
 & \ge -\sum_{i=1}^r \*q_i \log(\*q_i) + \sum_{i=1}^r \*q_i  \sum_{j=1}^m \theta_j \log(\*A_{:i}^\top  \*Q^j \*e) \nonumber
\end{align}
Here we denoted by $\*A_{:i}$, the $i^\text{th}$ column of $\*A$.
In the last equation, we used concavity of $log(\theta)$ and the result follows from Jensen's inequality. This shows that $-\kl(\*q|\*p)$ is an upper bound on the expression on the last line. The expression on the last line represents the expected negative KL divergence realized by a policy that we ideally  should be maximizing or at least a lower bound on it. However, if we maximize  $-\kl(\*q|\*p)$, we actually maximize an upper bound on what we really care about optimizing and this does not guarantee anything for the quantity we really care about. We avoid this problem completely in ~\eqref{eq:excalibr-abs} by considering linear deviation between the baseline and predicted  category distributions.

\section{Efficient ExCalibR}
\label{sec:efficiency}
We considered a full set of $n$ items and $n$ slots in the above section. For computational efficiency reasons, it is better to consider only a small number $k$ of slots for ranking. With a smaller $k$, we make a few changes in our setup for efficiency. Typically in a ranking problem, the top few slots get the most attention from users. 

First, we note that in the definition of $\*e$, we set $\*e_j=0$ for $ k \le j \le n$. Next we take a close look at the ExCalibR formulation \eqref{eq:excalibr-abs}. $\*e$ interacts with the matrix $\*P$, since the values in the last $n-k$ columns of $\*P$ do not contribute any scores in this setup, we can simply exclude them from the optimization. However, the last constraint ensures that $\*P$ is a doubly stochastic matrix. We still require that each of the first $k$ columns sum to 1. Now that we exclude some of the columns, the rows are required to sum to less than 1. We denote by $\hat{\*e} \in \real^k$  the vector that has only non-zero values. Denoting by $\hat{\*P} \in  \real^{n \times k}$, we solve the following optimization problem instead of formulation \eqref{eq:excalibr-abs}:

\begin{align}
\label{eq:excalibr-reduced}
\max_{\hat{\*P}, \*{\epsilon}}~ & (1 - \lambda) \mathbf{s}^\top \hat{\*P} \hat{\*e} - \lambda \epsilon^\top \*1 \\
\text{s.t.}~~& \*A^\top \hat{\*P} \hat{\*e} - \*q \le \epsilon  \nonumber \\ \nonumber
&  \*q - \*A^\top \hat{\*P} \hat{\*e} \le \epsilon  \\ \nonumber
& \hat{\*P} \*1 \le \*1~~~  \hat{\*P}^\top \*1 = \*1~~~  \hat{\*P} \ge \*0.
\end{align}

When we solve the above optimization, if any of the rows of $\hat{\*P}$ has only zeros, that means the item is never going to be placed in the top $k$ slots of the ranking. We eliminate all such items and consider only $m \le n$ items that have non-zero probability of being placed in the top $k$ positions for the Birkhoff-Von-Neumann decomposition. We overload the notation and use $\hat{\*P}$ to denote the matrix with any zero rows removed.

\begin{algorithm}[thp]
\caption{AugmentAndGetDS($\hat{\*P}$)}
\label{alg:augment}
\begin{algorithmic}[1]
    \State Input: $ \hat{\*P}\in \real^{m \times k}$ such that, $k < m$,~ $\hat{\*P}_{ij} \ge 0$ ,  $\hat{\*P}\*1 \le \*1$, $\*1^\top \hat{\*P} = \*1^\top$ \nonumber
    \State Output: $\*P \in \mathcal {DS}(m) $  \nonumber
    \State  $\*P \leftarrow \hat{\*P}$ 
    \State $\*x = \*1 - \hat{\*P}\*1$ \text{~~~~~/*remaining value to be filled per row*/}
    \While {$(\*x^\top \*1 > 0)$}
    \State $\*v \leftarrow \*0$ \text{~~~~~~/*initialize a $m$ dimensional vector */}
    \While {$\*v^\top 1 < 1$}
     \State $i^* \leftarrow \arg \max_{i} {\*x_i}_{i=1}^m$
     \If {$\*x_{i^*} + \*v^\top \*1 < 1$}
      \State \text{~~~~~/*Can add $\*x_{i^*}$ without sum of $v$ exceeding 1.*/}
      \State $\*v \leftarrow \*v + \*I_{i^*} \*x_{i^*}$ 
      \State \text{/* $\*I_{i^*}$ is a vector whose $i^*$th element is 1, others 0.*/}
      \State $\*x \leftarrow \*x - \*I_{i^*} \*x_{i^*}$
     \Else
      \State \text{~~~~~/*Add enough value to $\*v$ to make it sum to 1.*/}
      \State $\*v \leftarrow \*v  + \*I_{i^*} (1 - \*v^\top \*1)  $
      \State $\*x \leftarrow \*x  - \*I_{i^*} (1 - \*v^\top \*1)  $
      
      \EndIf
    \EndWhile
    \State $\*P.appendColumn(\*v)$
    \EndWhile
    \State {\bf return} $\*P$
\end{algorithmic}
\end{algorithm}

Once we solve for $\hat{\*P}$, we still need to construct a doubly stochastic matrix to be able to apply Birkhoff-Von Neumann decomposition which was originally defined on doubly stochastic matrices. We use the augmentation procedure shown in Algorithm \ref{alg:augment}  to obtain a doubly stochastic matrix with a small number of non-zero elements in the last $m-k$ columns. %Typically, the fewer the non-zero values in a doubly stochastic matrix, the fewer will be the permutation matrices we get from the BVN decomposition. 

The main idea of Algorithm \ref{alg:augment} is as follows. We first compute ($\*x$) how much value we should add in each row to make each row to add to one. We initialize the next column to add ($\*v$) to a vector of zeros. We pick the maximum element from $\*x$. We add this element to $\*v$ (but at the same row position) if the sum of elements of $\*v$ does exceed 1. If adding this element makes the sum of the column more than one, we just add enough to make the column sum one. We remove the added element value from $\*x$. We continue to follow this procedure until $\*x$ becomes zero. It is easy to see that this augmentation procedure gives a valid doubly stochastic matrix while satisfying constraints from the partial matrix learned in \eqref{eq:excalibr-reduced}.

\begin{algorithm}[thp]
\caption{Full steps for the computationally efficient procedure.}
\label{alg:full-steps}
\begin{algorithmic}[1]
\State Solve  \eqref{eq:excalibr-reduced}, obtain $\hat{\*P} \in \real^{n \times k}$ satisfying the constraints in \eqref{eq:excalibr-reduced}.
\State From $\hat{\*P}$, remove any rows with only zeros, get $\hat{\*P} \in \real^{m \times k}$. 
\State $\*P \leftarrow $ AugmentAndGetDS($\hat{\*P}$) 
\State Apply $BVN(\*P)$ on $m$ items that were retained in Step 2.
\end{algorithmic}
\end{algorithm}
Complete steps for the efficient version of our formulation are shown in Algorithm~\ref{alg:full-steps}. Simplifications of the optimization similar to that shown in our formulation (\ref{eq:excalibr-reduced}) were also considered in \cite{Rijke2022}. They also proposed an efficient algorithm to solve the resulting Birkhoff-von Neumann decomposition, while a comparison between our improvements and theirs would indeed be an interesting academic comparison, in our experiment we restrict ourselves to the core hypothesis of the paper and also show that most of the run-time gains actually come from the improvements in run-time of the LP (\ref{eq:excalibr-reduced}) rather than from the BVN decomposition steps. 

\begin{figure*}[thp!]
\begin{center}
\includegraphics[width=2.2in]{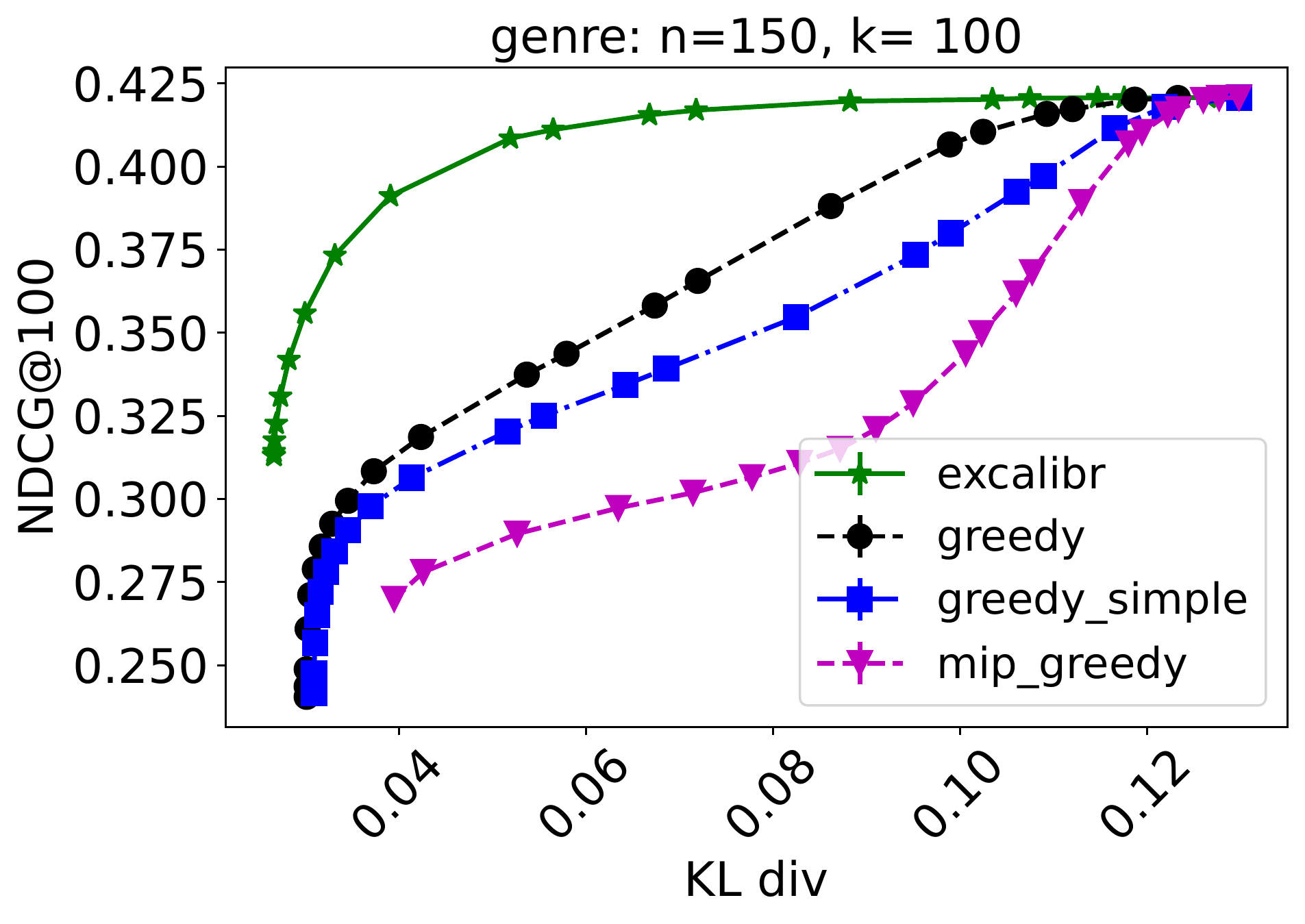}
\includegraphics[width=2.2in]{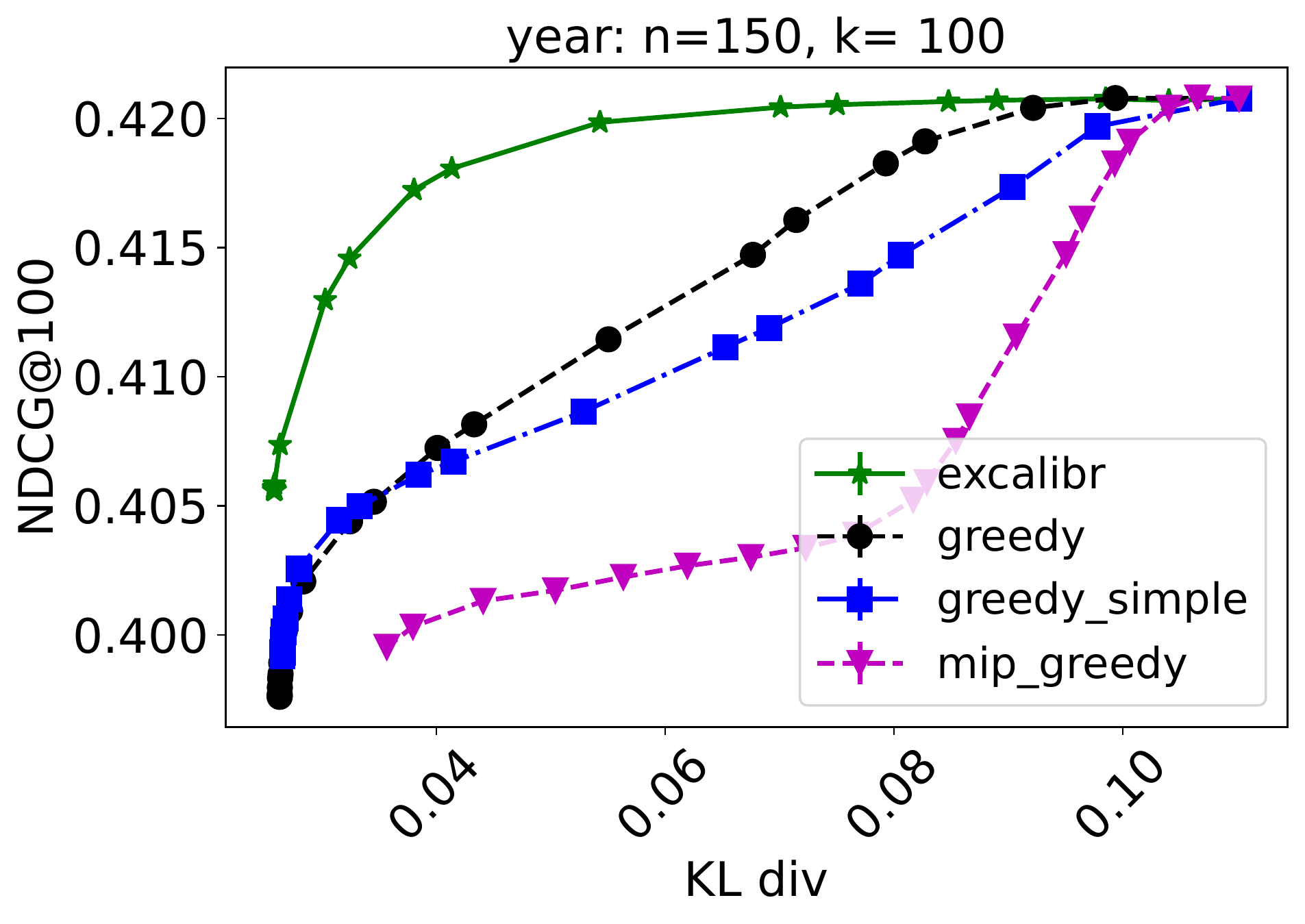}
\includegraphics[width=2.2in]{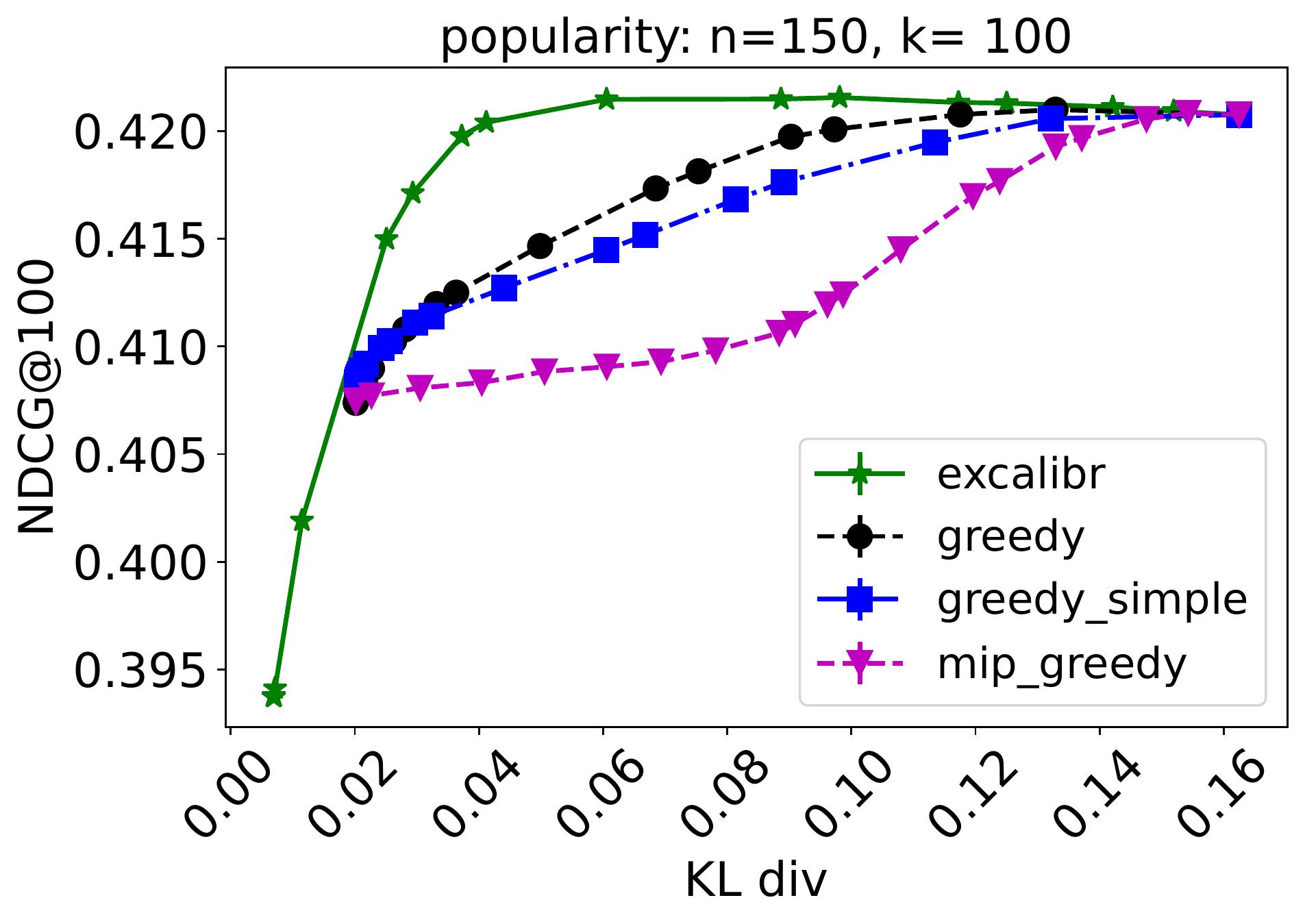} \\
\includegraphics[width=2.2in]{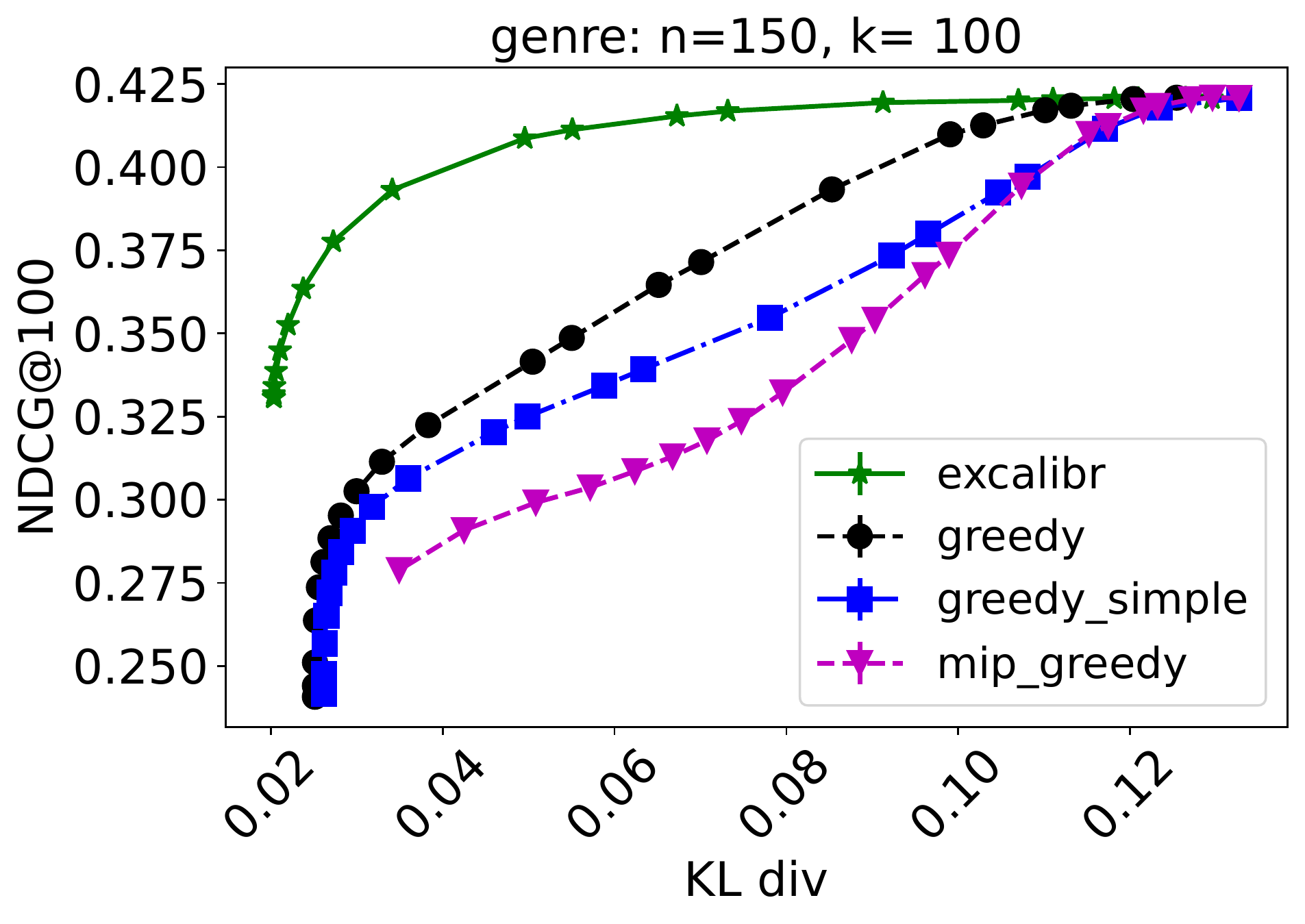}
\includegraphics[width=2.2in]{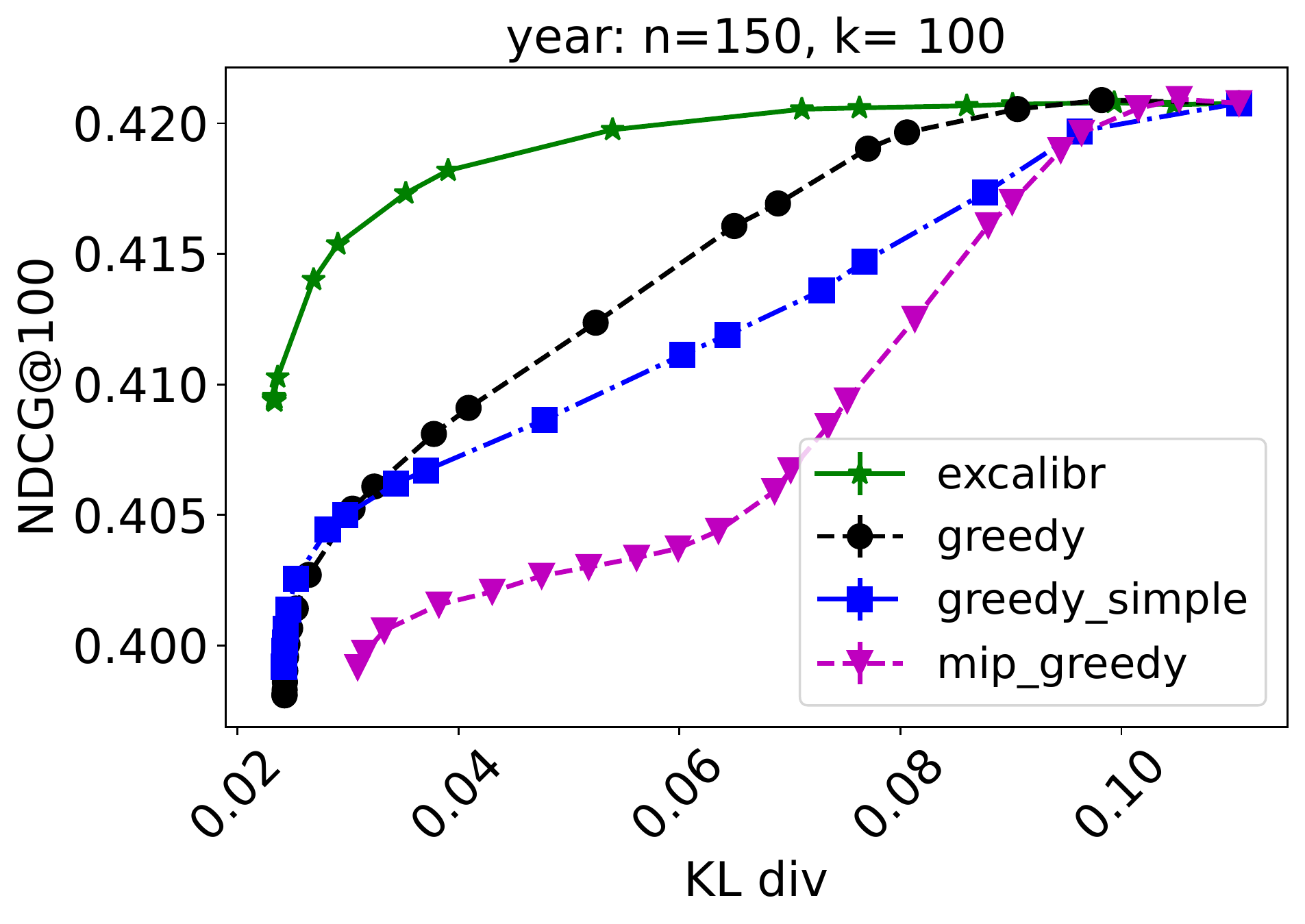}
\includegraphics[width=2.2in]{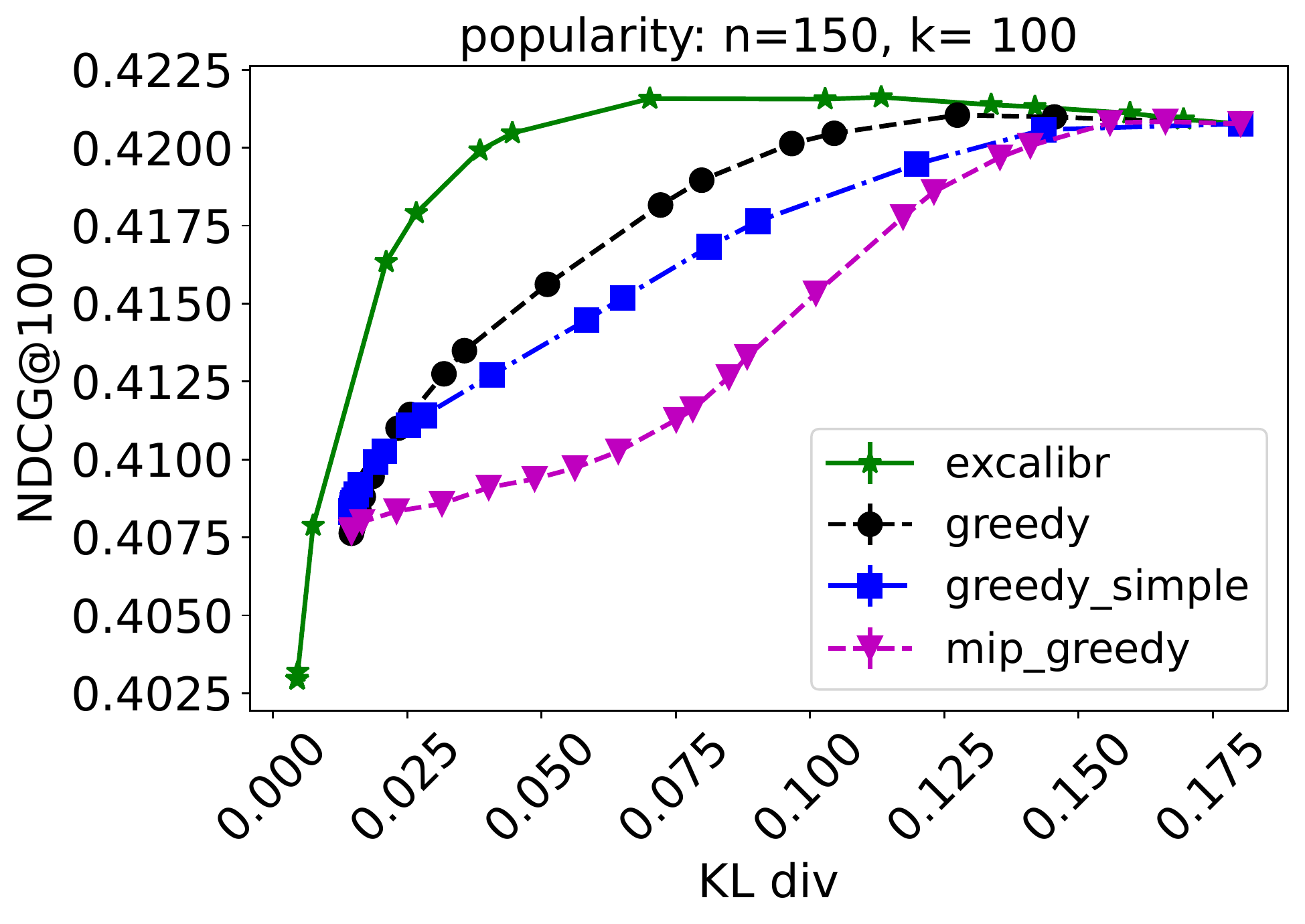} \\
 \vskip -0.05in
  \vskip -0.05in
  \caption{Top row: Results with positional weights determined by $\*e_i \propto \frac{1}{\log(i+1)}$, Bottom row: Results with positional weights determined by $\*e_i \propto \frac{1}{\sqrt{i}}$. These results are from $n=150$ and $k=100$. The curves show  NDCG at $100$ and KL divergence between user's past  history  distribution over genres, bucketized years and popularity levels obtained by top $k$ ranking respectively from left to right. {\em The {\bf standard errors} on the NDCG were so tiny that most visually clear differences are statistically significant.} The figures show that as the trade-off shifts towards calibration (KL divergence close to zero) from relevance (KL divergence maximum), ExCalibR performs much better compared to all baselines. Thus, for any non-trivial trade-off between relevance and KL divergence, ExCalibR gets a much superior NDCG for a given KL or a much smaller KL for a given NDCG. In the case of popularity, the NDCG even improves with calibration for ExCalibR. For the other two calibration categories, the plots show that it is possible to achieve a very similar NDCG  compared to the right most point with no calibration while being much more calibrated across all the three categories considered.}
  \vskip -0.15in
\label{fig:differences}
\end{center}
\end{figure*}

\begin{figure*}[thp!]
\begin{center}
\includegraphics[width=2.2in]{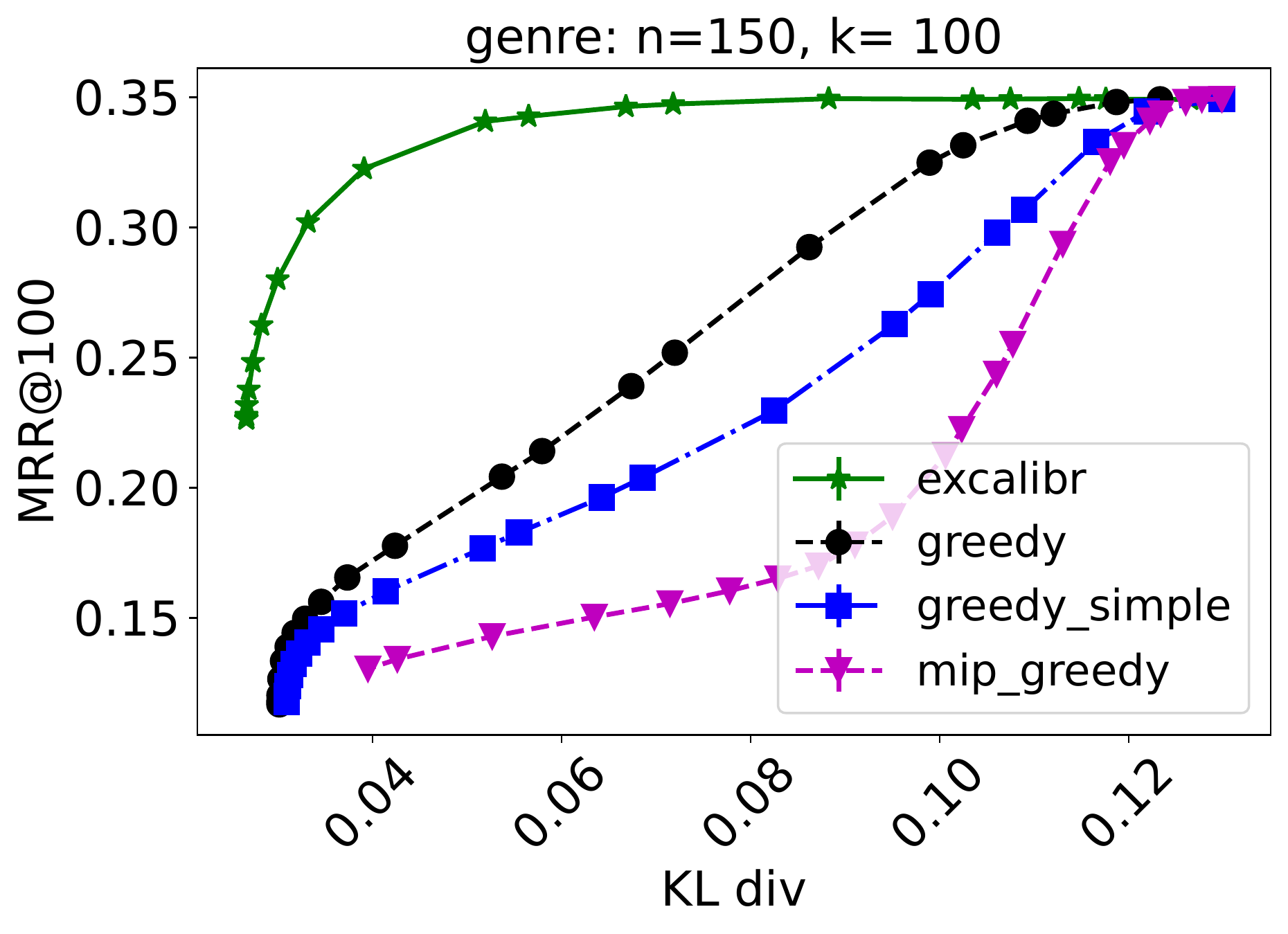}
\includegraphics[width=2.2in]{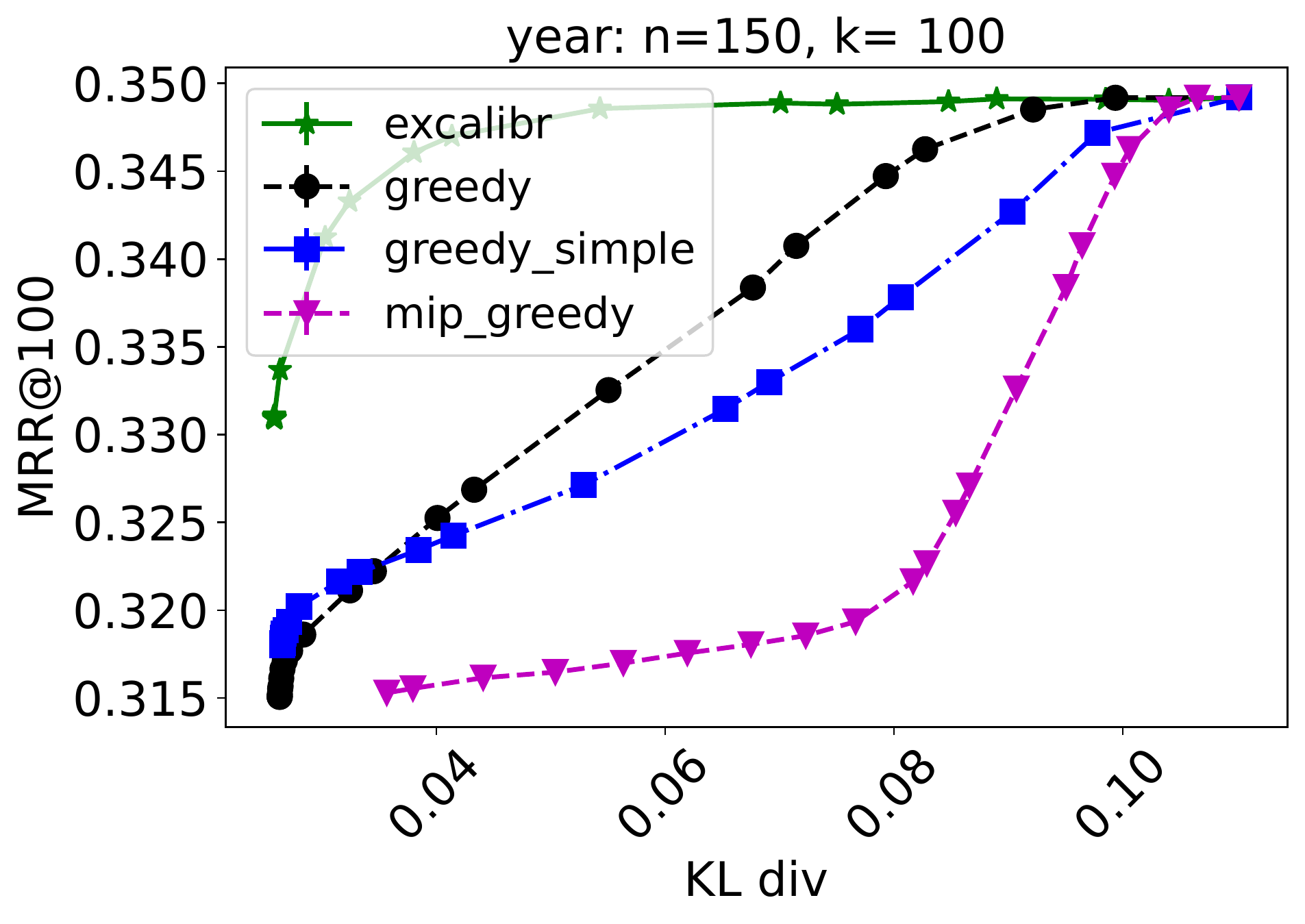}
\includegraphics[width=2.2in]{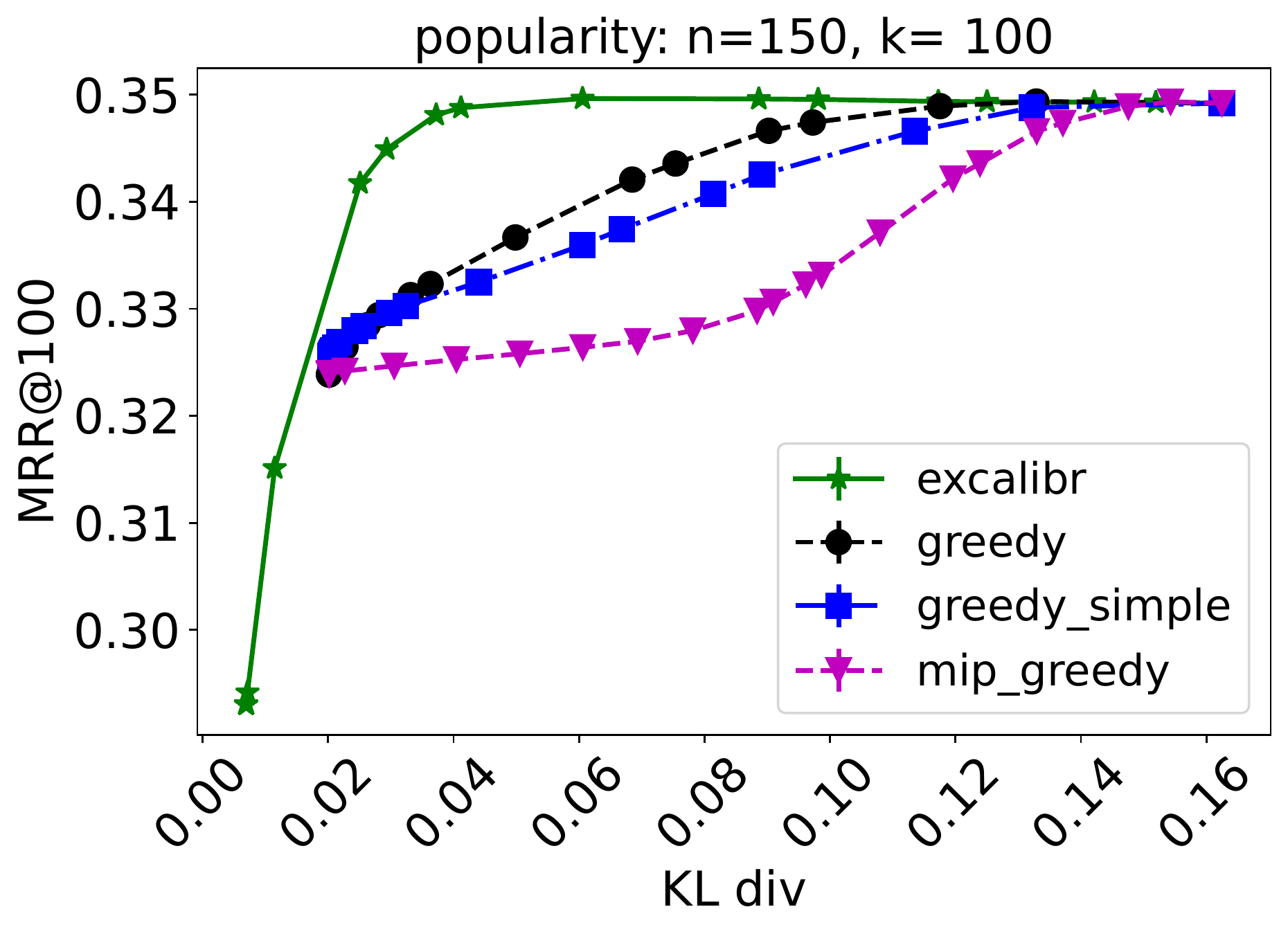} 
 \vskip -0.05in
  \vskip -0.05in
  \caption{Results from the same experiment as the top row in Figure~\ref{fig:differences} but we report MRR on the y-axis rather than NDCG. The MRR results were in line with NDCG results in all the experiments except for the numerical differences.}
  \vskip -0.15in
\label{fig:mrr-metric}
\end{center}
\end{figure*}

\begin{figure*}[thp!]
\begin{center}
\includegraphics[width=2.2in]{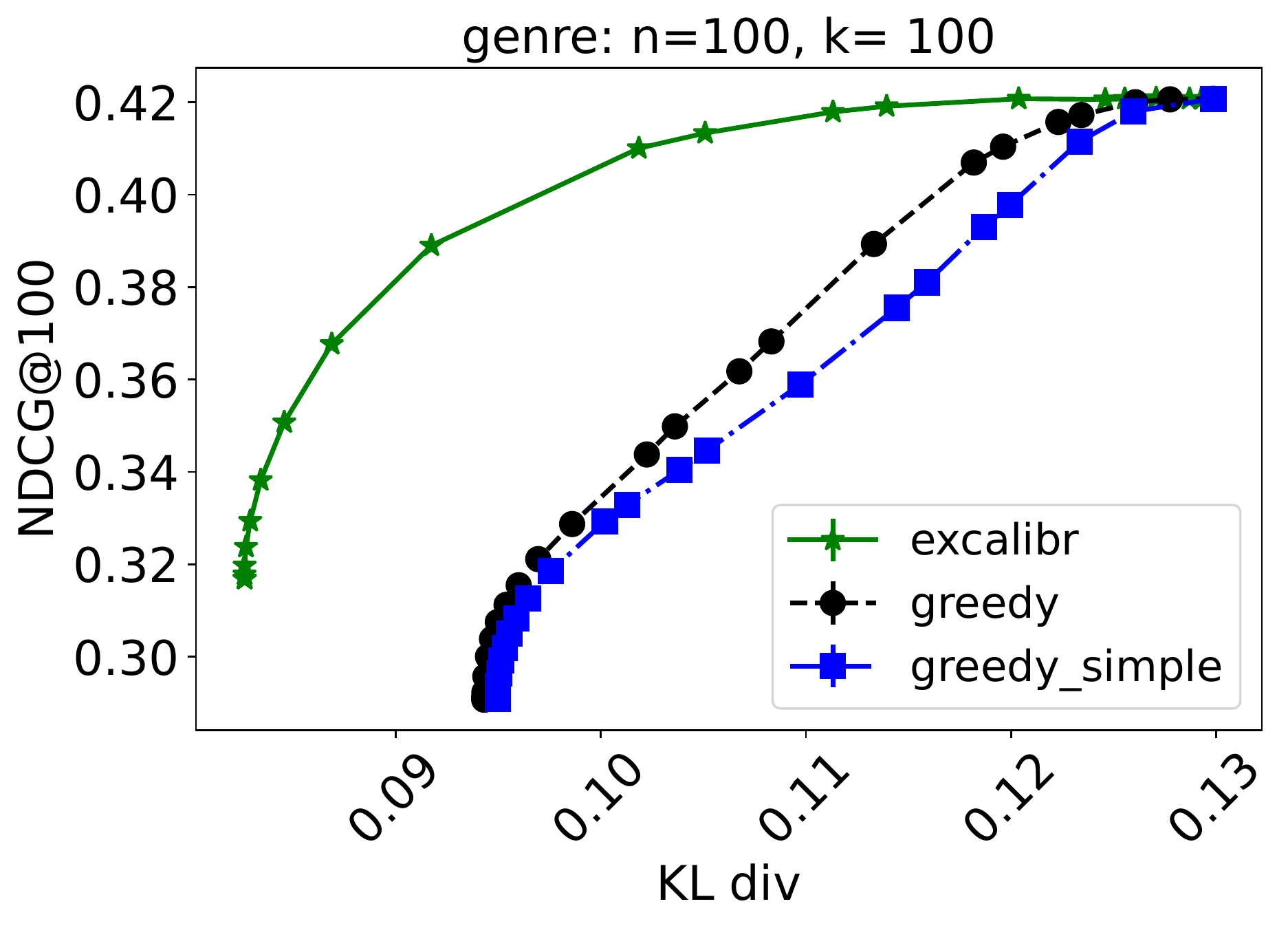}
\includegraphics[width=2.2in]{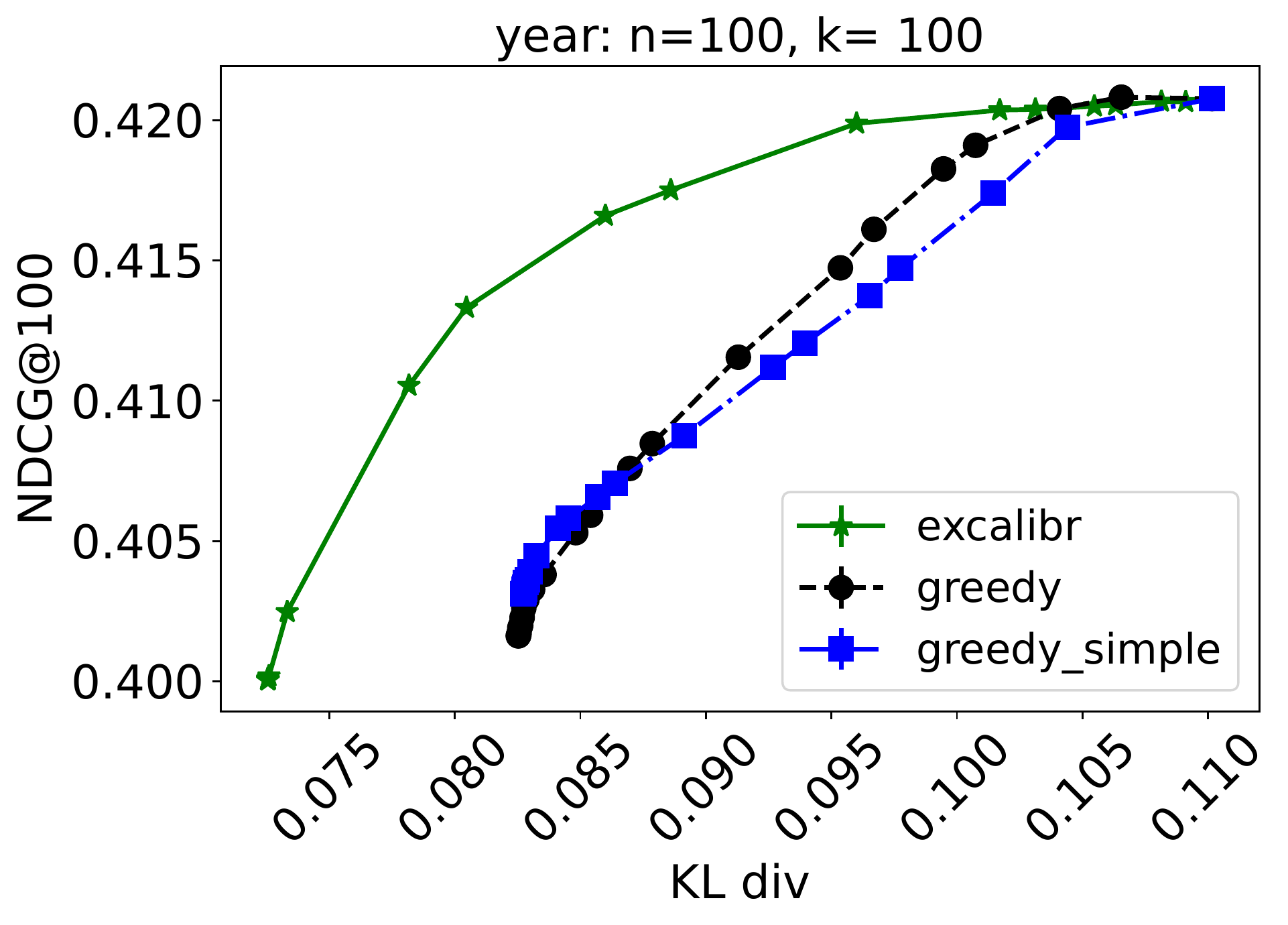}
\includegraphics[width=2.2in]{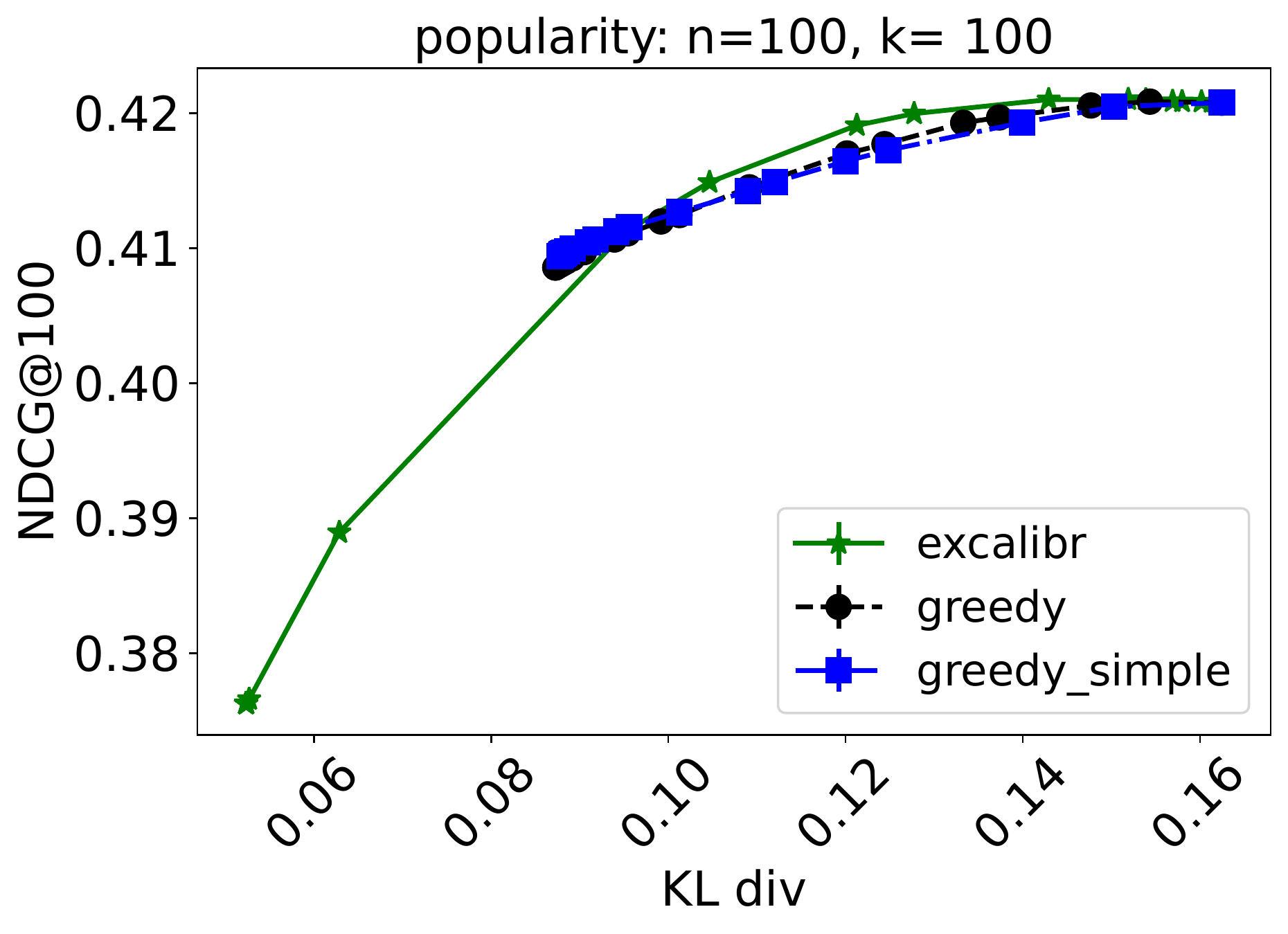} \\
\includegraphics[width=2.2in]{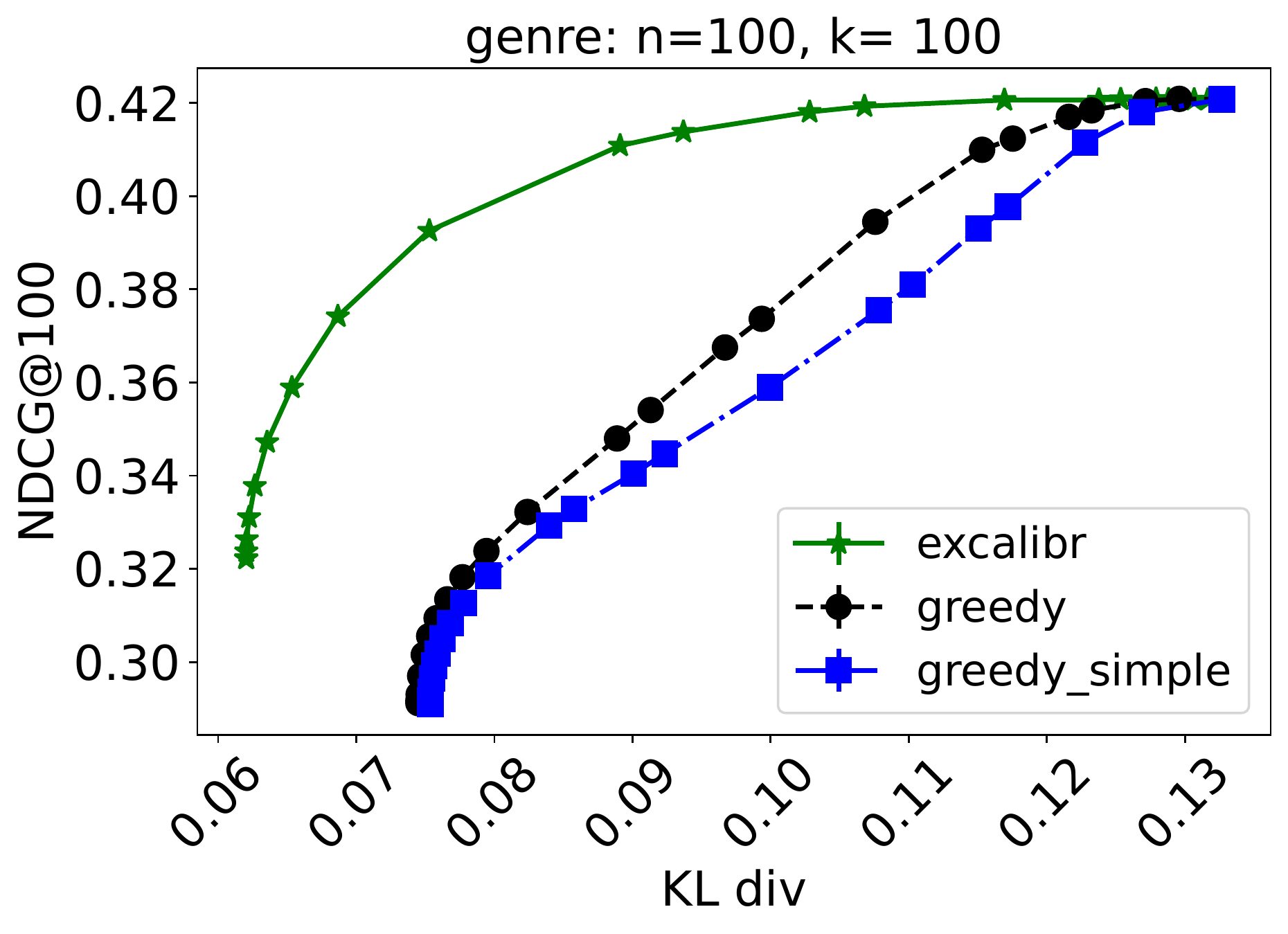}
\includegraphics[width=2.2in]{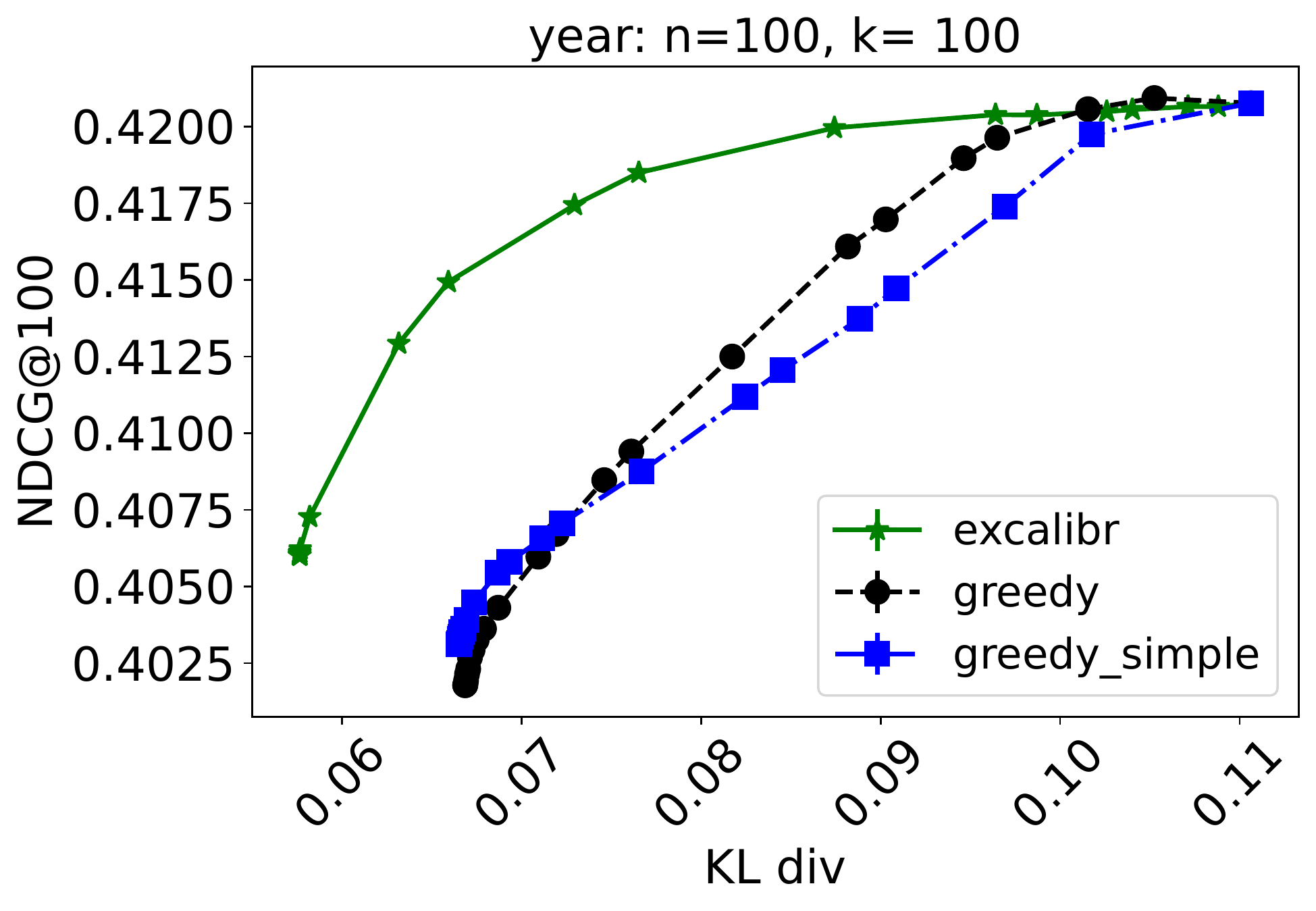}
\includegraphics[width=2.2in]{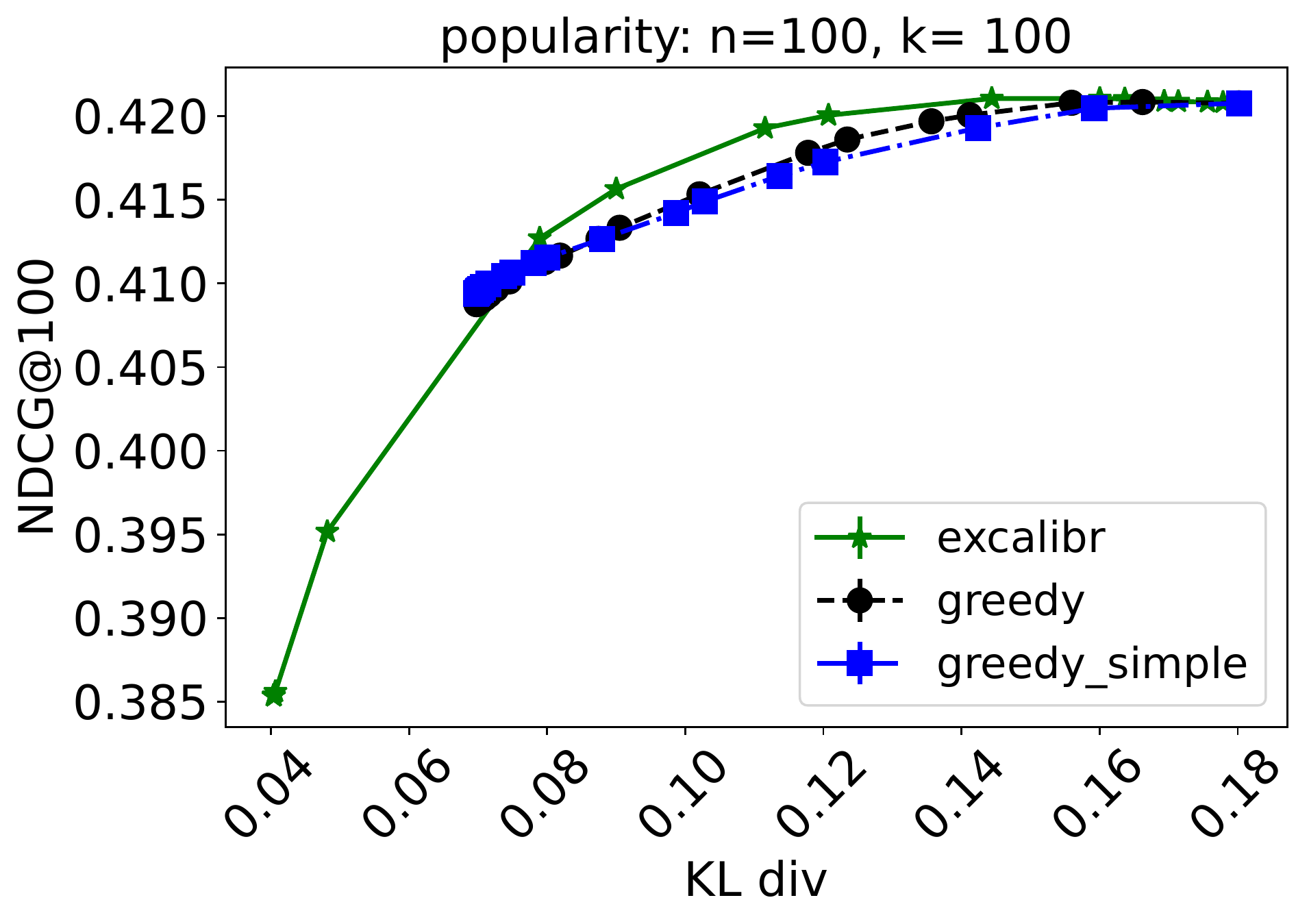} \\
 \vskip -0.05in
  \vskip -0.05in
  \caption{Top row: Results with positional weights determined by $\*e_i \propto \frac{1}{\log(i+1)}$, Bottom row: Results with positional weights determined by $\*e_i \propto \frac{1}{\sqrt{i}}$. These results are from $n=k=100$. The curves show  NDCG at $100$ and KL divergence between user's past  history  distribution over genres, bucketized years and popularity levels obtained by top $k$ ranking respectively from left to right. ExCalibR typically shows a significantly better trade-off compared to the baseline, but comparing these plots with the ones in~Figure \ref{fig:differences}, we see that using a larger $n$ compared to $k$ really helps much more for calibration.}
  \vskip -0.15in
\label{fig:rerank}
\end{center}
\end{figure*}

\begin{figure*}[thp!]
\begin{center}
\includegraphics[width=2.2in]{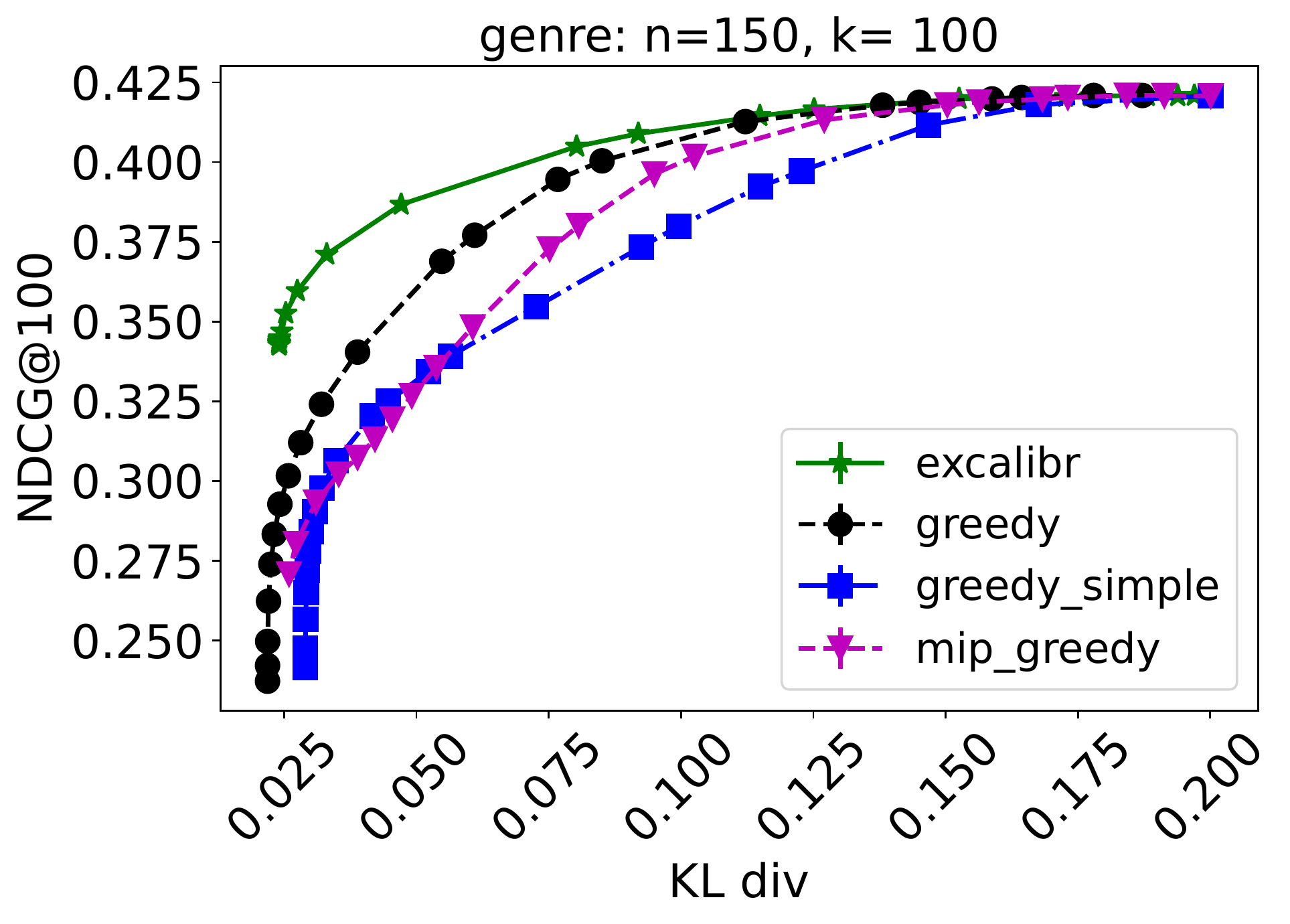}
\includegraphics[width=2.2in]{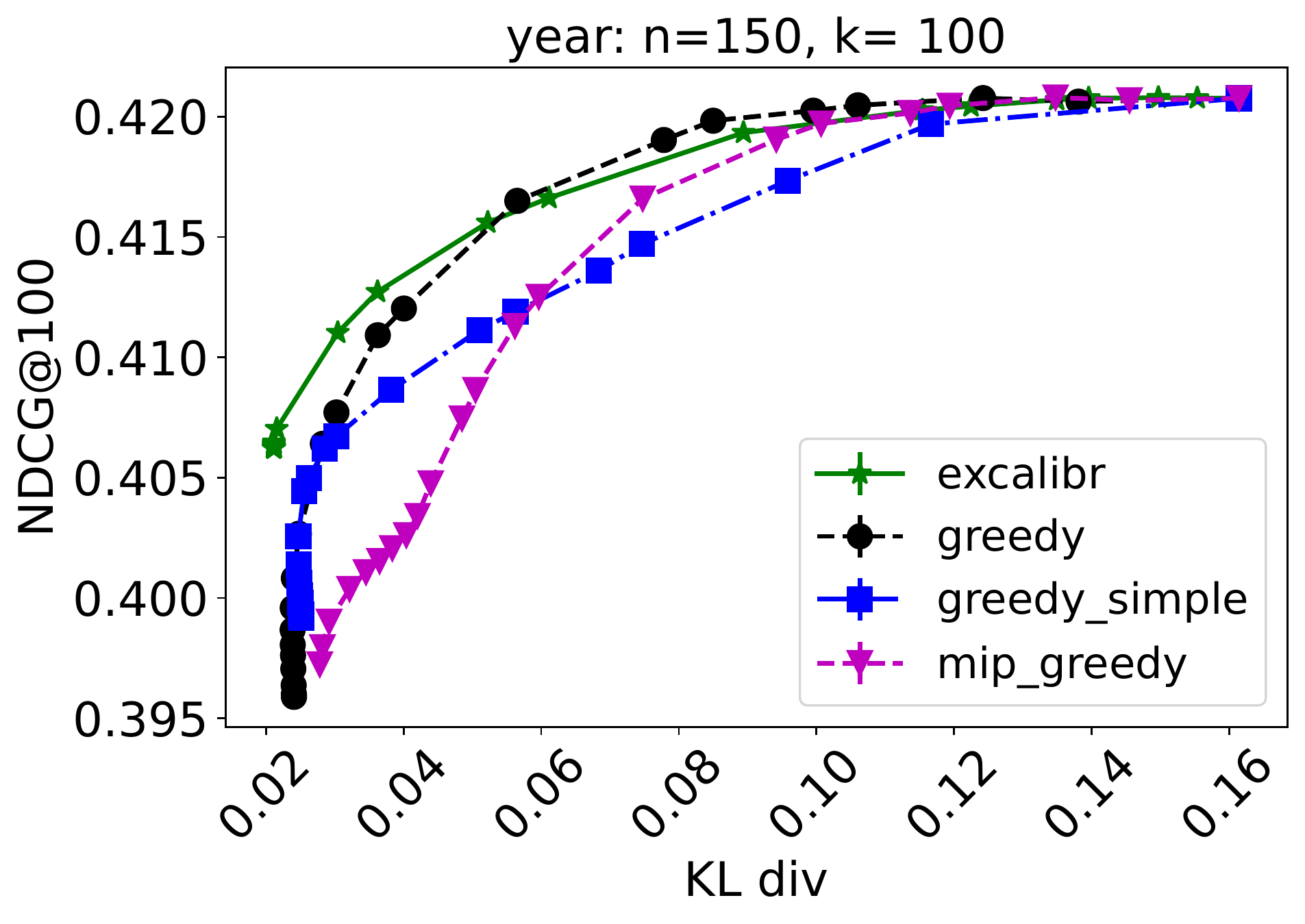}
\includegraphics[width=2.2in]{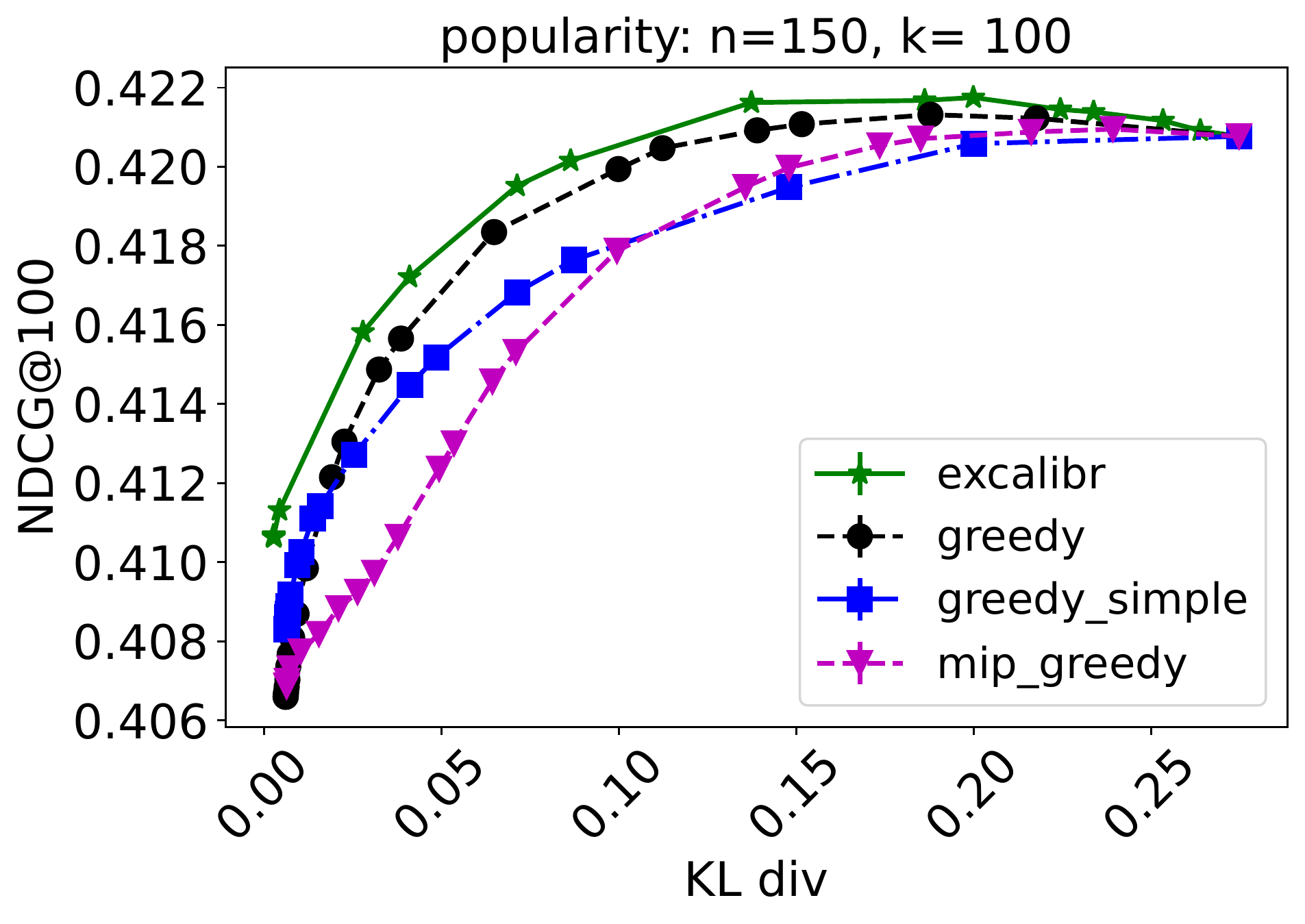} 
 \vskip -0.05in
  \vskip -0.05in
  \caption{ Results with positional weights determined by $\*e_i \propto \frac{1}{i}$ with $n=150$ and $k=100$. With significantly strong weights at the top positions with these position weights, the greedy algorithm manages to catch up with ExCalibR but is still not better than it.}
  \vskip -0.15in
\label{fig:mrr}
\end{center}
\end{figure*}

\begin{table}[]
    \centering
    \begin{tabular}{|l|c|c|c|}
    \hline
    Calibration Attribute & genre & bucketized year & popularity \\
    \hline
    Number of dimensions & 20 & 10 & 2  \\
    \hline
    Num. genres per item & 2.04 & 1 & 1 \\
    \hline
    Number of items  & \multicolumn{3}{c|}{20720}   \\
    \hline
    Num. train users & \multicolumn{3}{c|}{126k}  \\
    \hline
    Validation/Test users & \multicolumn{3}{c|}{5k} \\
    \hline
    \end{tabular}
    \caption{High level characteristics of the dataset along with the calibration attributes.}
    \vskip -0.2in
    \label{tab:datasets}
\end{table}

\section{Experiments}
In this section, we present experiments on a publicly available dataset to study the efficacy of the proposed method. 

\subsection{Dataset}

We used the movie lens 20 million dataset (ML20M) \cite{movielens}. Our implementations were done in python where we implemented the ExCalibR optimization using CVXPY \cite{cvxpy1,cvxpy2} using OR-Tools package \cite{ortools} for solving the LP optimization problem.

ML20M dataset has ratings for a sub-set of the items for a large number of users. We  removed any user that had fewer than five interactions in total.  We treated any rating above 3.5 as a positive interactions and all others as 0.  We then randomly divided users into train, validation and test users. Validation and test user data was further divided into two parts by randomly taking 80\% of the items with positive labels as user's past history of views and the remaining 20\% of the items with positive label as items to be predicted. This setup is similar to what has been used in previous papers \cite{Steck19,vae} for evaluating recommender models. We then trained an EASE model \cite{Steck19} with the hyper-parameters tuned for best NDCG performance on the held out validation set.  

The dataset also includes genre information for each movie where there could be multiple genres per movie. We gave an equal weight to all the listed genres to form the matrix $A$. We also considered year of each movie and considered a bucketized version in units of decades from 1920s to 2010s to generate another calibration attribute. For example, a movie from 1980s would have a calibration attribute indicating 1980s. We also considered the popularity as another category. We considered the top 5\% movies with maximum positives in the dataset as the popular titles and the remaining movies as the less popular titles. The top 5\% movies and the remaining movies had roughly the same number of positives. 

Some high level characteristics of the dataset are given in Table~\ref{tab:datasets}.  The held out 80\% user data was used for constructing a baseline distribution over genres, years or popularity based on the past viewing behavior of users. The goal was then to rank any title not in a user's past history such that we optimized the performance (in terms of relevance and calibration) on the held out items. We considered two different setups. In the first setup we chose n=150, k=100, this is a problem of top-k ranking from a larger set. We also considered another setup where n=100, k=100 which is simply a reranking problem.

\subsection{Baselines}
We used a number of baselines to compare our proposed technique against. We considered the  unweighted greedy directly from \cite{Steck19} as shown in \eqref{eq:kl-greedy} referred to as {\bf greedy simple} in the plots. The order of items selected by the greedy algorithm was used as the final ranking order. We also considered the modified  weighted greedy as shown in \eqref{eq:kl-greedy-weighted} referred to as {\bf greedy} . Obviously, the unweighted objective, deviates from the weighted objective in the ranking problem, we do not expect it to perform well but we included it for completeness.

Further, for the sake of being comprehensive, we also considered the Mixed Integer Program approach proposed in \cite{mip}. The basic idea in this approach is to select $k$ items out of $n$ using binary variables trading off an objective similar to ours \eqref{eq:excalibr-abs} but using a Mixed Integer Program. However, since this approach is for a sub-set selection problem rather than for rank ordering, we considered a simple heuristic on top of it to get a ranking from the selected sub-set of size $k$. After the sub-set selection with an unweighted objective to select $k$ items from the MIP, we give that set of items to the weighted greedy algorithm, we refer to this baseline as {\bf MIP greedy}.  Extending the MIP approach to a full weighted ranking algorithm might have solubility issues with $nxk$ binary variables and is beyond the scope of this paper. All mixed integer problems were solved using the package SCIP \cite{scip1,scip2} using its interface with CVXPY \cite{cvxpy1}. Note that these MIP procedures are applicable only when $n > k$ since sub-set selection makes no sense when $n=k$.

\subsection{How does ExCalibR perform compared to other baselines?}
We now compare the efficient ExCalibR (\ref{eq:excalibr-reduced}) with greedy optimization from \eqref{eq:kl-greedy-weighted} along with other baselines.
For the experiments in this section, we set $\*e_i \propto \frac{1}{\log(i+1)}$ and also $\*e_i \propto \frac{1}{\sqrt{i}}$.  All the optimization problems  have a trade-off parameter $\lambda$ between the relevance score and how close the two distributions (baseline category distribution and a  category distribution induced by a ranking) are. However,  since the different approaches are based on different objectives (for example, the greedy approach trades-off a relevance score with KL-divergence, ExCalibR trades-off between a relevance score and a linear deviation between the distributions), we cannot directly compare the results from the two methods for a single value of the trade-off parameter $\lambda$ which could represent a very different underlying trade-off.  Therefore, we swept over values of trade-off parameters in all the cases over a range of values starting from full weight on relevance to decreasing values on relevance. In each case, for both the algorithms, we computed the ranking and then computed NDCG@k when optimizing with $k$ position weights. We then looked at the entire trade-off curve comparing KL-divergence on one axis and NDCG@k on the other axis with $\lambda$ ranging from a value close to 1 a value close to 0.  In all our experiments, based on the EASE model score for items, we first select the top $n$ items for each user. We then use either the greedy procedure or ExCalibR to get top $k$ items among them with the specified $\*e$ value described above. 

The  results are shown in Figure~\ref{fig:differences} for three categories: genres, bucketized years and popularity. We note that ExCalibR is able to achieve a much superior trade-off compared to all the baselines in most settings. Typically, we do not want to deviate too far from the optimal NDCG that can be achieved in a ranking system. So the most interesting areas in the plots are the non-extreme areas on the x-axis. It can be seen that for a tiny drop or no drop (and even a little gain in the case of popularity) in NDCG compared to the right most point, a significantly superior calibration can be achieved by ExCalibR. The {\bf error bars} in these experiments were so tiny that they were almost not visible in the plots hence most of the visual differences are significant.

\subsection{What happens if we change the metric to MRR instead of NDCG?}
We also looked at the Mean Reciprocal Rank metric instead of NDCG for many of our experiments. The overall trends were very similar compared to the NDCG results in all experiments. We show the results from one such experiment in Figure~\ref{fig:mrr-metric}. These results can be compared with the top row of Figure~\ref{fig:differences} to see that the trends are exactly similar except for the numerical differences.

\subsection{Did $n>k$ in the previous experiment help?}
In the experiments so far, we performed top-k ranking with $n=150$ and $k=100$. In this section, we show  results on a reranking task with $n=k=100.$ The results are shown in Figure~\ref{fig:rerank}. In this section we exclude MIP greedy since it is based on a set selection algorithm. Selecting a set of size $k=100$ from $n=100$ would give the same set and moreover, MIP greedy was shown not to perform well already in the previous section.

From the results in Figure~\ref{fig:rerank} we see that ExCalibR continues to have advantage over the greedy baselines but compared to the case $n=150$ and $k=100$ the increases are lower and also that the ExCalibR curve drops of earlier than before as we calibrate more. This shows the value of performing a ranking over top $k$ slots from a larger set of $n$ items. We also tried $n=200$ and $k=100$ which did not show much of a difference compared to the case $n=150$ and $k=100$.

\subsection{What happens when we change the position weights to be much more top heavy?}
In this experiment, we modified the position weights. Rather than $\frac{1}{\log(i+1)}$ and $\frac{1}{\sqrt{i}}$ that were used so far, we now used a much more aggressive $\*e_i \propto \frac{1}{i}$. Since it becomes much more top heavy, the greedy algorithm bridges some of the gaps from the earlier experiments. Even though it doesn't convincingly outperform ExCalibR, it is nearly on par with it for the most part. This can be seen in Figure~\ref{fig:mrr}.
To give an idea of how these three weightings differ, with normalized weights over 100 slots, $\frac{1}{\log(i+1)}$ is around 14\% of the weight in the top 5 positions, around 21.6\% over the top 10. Similarly, $\frac{1}{\sqrt{i}}$ is around 17\% of the weight over the top 5 positions and 27\% over the top 10. Finally, $\frac{1}{i}$ is 44\% in the top 5 positions and 56\% in the top 10. With this heavy skew, the greedy algorithm can make good selections since it is essentially greedily picking up the best items for the initial slots.

\begin{table}[]
    \centering
    \begin{tabular}{|l|c|c|c|}
    \hline
    Method & genre & year & popularity \\
    \hline
    ExCalibR  & 0.69 $\pm$ 0.02 & 0.43 $\pm$ 0.01 & 0.39 $\pm$ 0.00 \\ 
    Greedy & 0.51 $\pm$ 0.00 & 0.50 $\pm$ 0.00  & 0.47  $\pm$ 0.00    \\
    MIP greedy & 0.37 $\pm$ 0.04 & 0.27 $\pm$ 0.01 & 0.28  $\pm$ 0.01   \\
    \hline
    \end{tabular}
    \caption{Runtime in second per user for the three approaches. ExCalibR takes the same order of time as greedy and is in fact faster than our greedy optimization implementation. The higher error bar for ExCalibR and MIP is due to the fact that it takes more time when the LP optimization is harder (i.e., higher weight on calibration).}
    \vskip -0.2in
    \label{tab:runtimes}
\end{table}

\subsection{How does the runtime of ExCalibR compare with other methods?}
Next, we compare the run-times of the greedy approach and the MIP approach with that of ExCalibR. Most of the setup was similar to the description in the previous section.  We report the average run-time for ExCalibR, greedy optimization and the MIP approach per user. The results are shown in Table~\ref{tab:runtimes} by avergaing the runtime over the runs for a range of trade-off values. We notice that the ExCalibR approach takes either slightly more time compared to greedy in the case of genre and even less time in the case of years and popularity. The greedy approach has to go over the entire remaining candidate set of up to 150 items at each position repeatedly computing the objective during each attempt. For ExCalibR, if we further break down the total time taken by ExCalibR into time taken by the Linear Program and for BVN decomposition, we saw that most of the time was taken by the Linear Program itself.  For instance, for genre, of the 0.69 seconds, the LP took 0.648 seconds and BVN decomposition under 0.05 seconds. For year, the LP took $0.40$ while BVN decomposition took $0.025$ seconds. For popularity, the LP took $0.39$ and BVN decomposition under 0.01  seconds. Using commercial solvers or specialized solvers may be useful for speeding up the ExCalibR approach.

\begin{table}[]
    \centering
    \begin{tabular}{|l|c|c|c|}
    \hline
    Method & genre & year & popularity \\
    \hline
    Total time  & 33.61\%  & 34.97\%  &  33.61\% \\ 
    LP time &  32.51\% & 31.62\% &   32.51\%    \\
    BVN time & 43.59\%  & 58.53\%  & 43.35\%\\
    \hline
    \end{tabular}
    \caption{The percentage drop in runtime between an optimized version of ExCalibR from an unoptimized version. }
    \vskip -0.2in
    \label{tab:optimization}
\end{table}

\subsection{How much did the optimizations help?}
We now show how much the optimizations from Section~\ref{sec:efficiency} helped in reducing the run-time. 
For n=150 and k=100 we solved ExCalibR with all the optimizations from Section~\ref{sec:efficiency}  and compared it against a version with no optimizations but by setting $\*e_i=0$ for $i > k$ and then solving the full $n \times n$ optimization problem.  The results from this experiment are shown in Table~\ref{tab:optimization}. It can be seen that there is more than 33\% drop in the overall time taken.

\section{Conclusions}
In this paper, we studied the problem of achieving calibration trading it off with relevance over any set of given attributes. We set up the problem of calibration in such a way that the distribution over calibration attributes becomes a linear transformation of known quantities and an unknown doubly stochastic matrix that we learned using linear program. The learned doubly stochastic matrix was decomposed via Birkhoff-von Neumann decomposition can achieve much superior trade-off compared to many other baselines. While our formulations can potentially be computationally expensive, it was shown that many optimizations can be considered such that the resulting problem can still be solved in time frames comparable to that of a greedy approach. We considered one formulation in this paper, many variants of the formulation including bounding the relative differences, considering other types of objectives etc. might be possible. It is also an interesting direction to consider ExCalibR in depth for other identified problems such as calibration for intent priors, calibration for coverage of producers in an e-commerce platform etc. There is also potential for further runtime improvements from using commercial LP solvers or by writing custom optimization solutions for ExCalibR LP. Finally, we consider the problem along a single dimensions, it is possible to consider multiple calibration attributes simultaneously which may be much more suitable in many applications.

%\Appendix

%Initially $\*x^\top \*1 = m-k$ This is because each columns of $\*P$ add to 1.  

%We reduce $\*x^\top \*1$ by 1 in each inner loop.

%All rows of $P$ we get from Augment add to 1.
%Whenever we pick an element from $\*x$ to add, we add it to the same row. This is evident from $\*I_{j^*}$ which is one at the same index where the maximum was found.

%When we exit the inner while loop, we form a column that adds to one. This is easy to see because we keep adding a number to the column as long as it less than 1 or if becomes 1 when we add, we add just enough to the column such that the sum of the elements in the column is one. 

\section{Acknowledgements}
The author would like to thank Harald Steck for many helpful discussions and his help with the dataset used in this paper.

\bibliographystyle{ACM-Reference-Format}
\bibliography{main}

\end{document}